\documentclass[twocolumn]{aastex631}

\usepackage{slantsc}
\usepackage{threeparttable}
\usepackage{amsmath}
\usepackage{placeins}

\begin{document}

\title{High Resolution Spectroscopic Follow-up Observation Results for 13 EMP Candidates Selected by Narrow-band Photometry}

\author[0009-0008-2988-2680]{Huiling Chen}
\affiliation{Department of Astronomy, School of Physics, Peking University, Beijing 100871, China}
\affiliation{Kavli Institute for Astronomy and Astrophysics, Peking University, Beijing 100871, China}

\author[0000-0003-3250-2876]{Yang Huang}
\affiliation{School of Astronomy and Space Science, University of Chinese Academy of Science, Beijing 100049, China}
\affiliation{National Astronomical Observatories, Chinese Academy of Sciences, Beijing, 100101, China}

\author[0000-0002-7727-1699]{Huawei Zhang}
\affiliation{Department of Astronomy, School of Physics, Peking University, Beijing 100871, China}
\affiliation{Kavli Institute for Astronomy and Astrophysics, Peking University, Beijing 100871, China}

\author[0000-0003-0292-4832]{Zhen Guo}
\affiliation{Instituto de Física y Astronomía, Universidad de Valparaíso, ave. Gran Bretaña, 1111, Casilla 5030, Valparaíso, Chile}
\affiliation{Chinese Academy of Sciences South America Center for Astronomy (CASSACA), National Astronomical Observatories, CAS, Beijing, 100101, China}

\correspondingauthor{Yang Huang and Huawei Zhang}
\email{huangyang@ucas.ac.cn and zhanghw@pku.edu.cn}

\begin{abstract}

Extremely metal-poor (EMP; $\mathrm{[Fe/H]} < -3.0$) stars preserve key information about the earliest stages of Galactic chemical evolution. Based on a training set of spectroscopic metallicities, we constructed catalogs of more than 120,000 EMP candidates from large-scale narrow-band photometric surveys. To validate this approach, we obtained high-resolution CFHT/ESPaDOnS spectra for 13 candidates selected from the SkyMapper-based catalog. The follow-up observation confirms that the narrow-band photometric selection is effective in identifying very metal-poor stars and retaining a substantial EMP fraction: four targets are confirmed as EMP stars, and all targets remain very metal-poor with $\mathrm{[Fe/H]} < -2.4$. The photometric metallicities are systematically lower than the spectroscopic values by 0.44 dex, with a relatively small scatter of 0.19 dex, indicating a systematic offset in the photometric metallicity scale that could be reduced through improved calibration. Abundances for more than 20 chemical species are derived, leading to the identification of a new potassium-enhanced r-II star, J2001-1215. By combining chemical abundances with orbital properties, five stars are found to be dynamically consistent with known Galactic substructures, including Gaia--Sausage--Enceladus, Thamnos, and Sequoia. These stars provide high-resolution chemical measurements for the extremely metal-poor regime of these Galactic substructures. These results underscore the role of narrow-band photometry in efficiently selecting EMP candidates for targeted high-resolution spectroscopic follow-up, thereby enabling detailed chemical and dynamical studies of the early Milky Way.

\end{abstract}


\section{Introduction}\label{sec:intro}

The first generation of Population III stars, formed from pristine gas, mark the beginning of the chemical enrichment history of the universe. None of these ancient stars have been directly observed, as they were likely very massive and short-lived \citep{2004ARA&A..42...79B}. As an alternative, low-metallicity stars such as very metal-poor (VMP; $\mathrm{[Fe/H]<-2.0}$) and extremely metal-poor (EMP; $\mathrm{[Fe/H]<-3.0}$) stars have become key targets \citep{2005ARA&A..43..531B,2015ARA&A..53..631F}. Formed from material enriched by the Population III stars, these low-mass stars could survive to the present day, preserving in their atmospheres the chemical signatures of their birth clouds. These signatures enable studies of the first stars' properties, nucleosynthesis processes, and the astrophysical sites responsible for producing various elements \citep{2018ARNPS..68..237F,2020ApJ...900..179K}.

Although VMP and EMP stars are valuable tracers of early stellar and Galactic evolution, they are intrinsically rare and therefore challenging to identify \citep[e.g.,][]{2025arXiv250514787K}. To expand the census of these rare objects, numerous large-scale sky surveys over the past decades have undertaken systematic searches for VMP and EMP candidates. Major spectroscopic efforts include the HK survey \citep{1985AJ.....90.2089B}, the Hamburg/ESO Survey (HES; \citealt{2006ApJ...652.1585F,2008A&A...484..721C}), SDSS \citep{2016A&A...593A..10A}, RAVE \citep{2018AJ....155..256P}, LAMOST \citep{2022ApJ...931..147L}, and {\it Gaia} \citep{2024A&A...683L..11V}. 
In parallel, narrow-band photometry targeting the Ca\,{\sc ii} H and K lines wavelength region has enabled efficient selection of metal-poor candidates over wide sky areas and to fainter magnitudes, as demonstrated by the SkyMapper Southern Survey \citep{2007PASA...24....1K}, the Pristine survey \citep{2017MNRAS.471.2587S}, and SAGES \citep{2018RAA....18..147Z, 2023ApJS..268....9F}. Multiple narrow-band photometric surveys can further provide estimates of carbon and $\alpha$-element abundances, as shown by J/S-PLUS \citep{2019A&A...622A.176C,2019MNRAS.489..241M}, enabling a more detailed characterization of metal-poor stellar populations across large samples.

Based on the latest released photometric catalogs \citep[e.g.,][]{2023ApJS..268....9F, 2024A&A...691A.138P, 2024PASA...41...61O}, we derived stellar atmospheric parameters for over 90 million stars, among which approximately 120,000 are classified as EMP candidates \citep{2019ApJS..243....7H,2022ApJ...925..164H,2023ApJ...957...65H,2024ApJ...974..192H,2025RNAAS...9...74H}. Our metallicity estimation method first constructs a training dataset using spectroscopic metallicity measurements from either high-resolution or low/medium-resolution surveys with stellar colors from narrow band surveys and {\it Gaia}. The relation between metallicity and color is then applied to photometric survey data. The typical uncertainty of metallicity derived by this method is 0.05--0.10\,dex for [Fe/H] $> -1.0$, increasing slightly toward lower metallicity, e.g., 0.20--0.30\,dex for [Fe/H] $< -2.0$ (see Fig.~12 in \citealt{2022ApJ...925..164H} for details). Within similar metallicity ranges, the method yields more accurate estimates for giants than for dwarfs. Overall, the precision of photometry-based metallicity estimates is comparable to, and in some cases even slightly better than, low-resolution spectroscopic measurements.

Beyond increasing the number of known VMP and EMP stars, such large photometric-metallicity catalogs also provide an important basis for studying the metal-poor components of Galactic substructures. Dynamically associated metal-poor stars in Galactic substructures preserve information about the progenitor systems from which they originated, including disrupted dwarf galaxies and globular clusters \citep[e.g.,][]{2018ApJ...862..114S,2020ApJ...891...39Y,2025ApJ...980...71V}. When combined with detailed chemical abundance measurements, these stars can be used to trace the early chemical evolution of different accreted systems and to constrain the assembly history of the Milky Way \citep[e.g.,][]{2022A&A...661A.103M}. However, VMP and especially EMP stars with high-resolution abundances remain sparsely sampled in known Galactic substructures. Expanding the spectroscopically confirmed sample of metal-poor stars selected from large photometric catalogs is therefore important not only for increasing the number of EMP stars, but also for identifying candidate members of Galactic substructures and characterizing their chemical properties.

To further evaluate the reliability of our EMP candidate catalog, expand the sample of spectroscopically confirmed EMP stars, and examine possible links between these metal-poor stars and known Galactic substructures, we are conducting follow-up observations with both medium- and high-resolution spectroscopy on various telescopes. Here, we present the first-stage high-resolution results for 13 EMP candidates selected from our catalog. The outline of the paper is as follows. Section\,\ref{Target Selection and Observation} describes the target selection and high-resolution observations. Section\,\ref{Spectrum Analysis of Program Stars} presents the determination of stellar parameters and elemental abundances. Section\,\ref{Kinematic} describes the derivation of kinematic and orbital parameters. Section\,\ref{Discussion} discusses the chemical signatures of individual stars and their possible associations with known Galactic substructures in a chemical-dynamical context. Finally, a summary is provided in Section\,\ref{sec:summary}.




\begin{table*}
\scriptsize
\setlength{\tabcolsep}{3.9pt}
\centering
\caption{
Observation details and basic photometric properties of the sample stars.
}
\begin{tabular}{cccrrcccrrcc}
\hline
\hline
ID & R.A. & Decl. & \multicolumn{1}{c}{$l$} & \multicolumn{1}{c}{$b$} & $E(B-V)$ & $\rm [Fe/H]_{phot}$ & Date & \multicolumn{1}{c}{$G$} & \multicolumn{1}{c}{Exposure Time} & S/N & S/N \\
 & (hms) & (dms) & \multicolumn{1}{c}{\makebox[0pt][c]{($^\circ$)}} & \multicolumn{1}{c}{\makebox[0pt][c]{($^\circ$)}} &  &  & (UT) & \multicolumn{1}{c}{\makebox[0pt][c]{(mag)}} & \multicolumn{1}{c}{\makebox[0pt][c]{(sec)}} & (4500\,\AA) & (5500\,\AA) \\
\hline

J0858-0809 & 08:58:05.90 & $-$08:09:17.0 & 236.2516 & 23.4406 & 0.027 & $-3.31$ & 2021-11-23 & 9.98 & 255$\times$1 & 30 & 58 \\
J1930-1739 & 19:30:08.80 & $-$17:39:17.8 & 21.0945 & $-16.3110$ & 0.105 & $-3.06$ & 2021-08-28 & 12.16 & 1900$\times$1 & 35 & 66 \\
J1944-0837 & 19:44:28.40 & $-$08:37:46.5 & 31.1378 & $-15.6778$ & 0.188 & $-3.37$ & 2021-08-28 & 12.37 & 1650$\times$2 & 28 & 73 \\
J1950-1255 & 19:50:36.50 & $-$12:55:28.5 & 27.7423 & $-18.8726$ & 0.122 & $-3.40$ & 2021-08-28 & 12.08 & 1800$\times$1 & 29 & 60 \\
J1952-1836 & 19:52:53.50 & $-$18:36:47.1 & 22.4195 & $-21.6540$ & 0.153 & $-3.26$ & 2021-08-28 & 11.42 & 920$\times$1 & 29 & 60 \\
J2001-1215 & 20:01:29.90 & $-$12:15:44.5 & 29.5590 & $-21.0132$ & 0.215 & $-3.12$ & 2021-08-28 & 11.43 & 960$\times$1 & 28 & 59 \\
J2020-1002 & 20:20:46.60 & $-$10:02:23.6 & 33.9052 & $-24.3635$ & 0.063 & $-3.08$ & 2021-11-28 & 12.10 & 1770$\times$1 & 26 & 54 \\
J2023-0815 & 20:23:44.10 & $-$08:15:48.4 & 36.0090 & $-24.2502$ & 0.048 & $-3.12$ & 2021-09-01 & 12.40 & 1680$\times$2 & 32 & 78 \\
J2026-0047 & 20:26:54.60 & $-$00:47:26.7 & 43.5974 & $-21.4923$ & 0.174 & $-3.19$ & 2021-08-28 & 11.99 & 1710$\times$1 & 35 & 63 \\
J2038-0023 & 20:38:43.20 & $-$00:23:33.3 & 45.5417 & $-23.8708$ & 0.068 & $-3.35$ & 2021-11-28 & 12.38 & 2380$\times$1 & 30 & 59 \\
J2043-1305 & 20:43:59.90 & $-$13:05:21.1 & 33.4599 & $-30.8041$ & 0.035 & $-3.12$ & 2021-09-01 & 11.71 & 1310$\times$1 & 43 & 69 \\
J2140-1227 & 21:40:23.00 & $-$12:27:04.4 & 41.3825 & $-43.0381$ & 0.048 & $-3.03$ & 2021-08-28 & 10.78 & 550$\times$1 & 36 & 64 \\
J2326-0159 & 23:26:53.00 & $-$01:59:25.2 & 80.4912 & $-57.7188$ & 0.047 & $-3.36$ & 2021-09-01 & 10.76 & 535$\times$1 & 32 & 64 \\

\hline
\end{tabular}
\label{tab:observation_and_photometry_details}
\end{table*}

\section{Target Selection and Observation} \label{Target Selection and Observation}

The EMP candidates are selected from our photometric metallicity catalog based on the SkyMapper Southern Survey (SMSS) DR2 data \citep{2022ApJ...925..164H}. Stellar atmospheric parameters in this catalog are derived using a training dataset calibrated with the LAMOST Galactic spectroscopic survey \citep{2012RAA....12..723Z,2012RAA....12.1243L} and the SDSS/SEGUE survey \citep{2009AJ....137.4377Y}. We first select stars with photometric metallicities of $\mathrm{[Fe/H]_{phot} < -3.0}$. To ensure the reliability of the candidate selection, we additionally require metallicity uncertainties smaller than 0.3\,dex. Possible unresolved binaries are excluded by imposing a {\it Gaia} RUWE threshold of $\mathrm{RUWE} < 1.4$. Because giant stars are intrinsically brighter and generally yield more robust photometric metallicity estimates, we further restrict the candidate sample to giant stars selected according to the criteria defined by \citet{2022ApJ...925..164H}.

To ensure observational efficiency and sufficiently high-quality spectra for detailed abundance analysis, we further restrict the candidate sample to bright giant stars with $G < 12.5$\,mag. From this selected sample, we chose targets observable during our allocated CFHT runs for high-resolution spectroscopic follow-up with ESPaDOnS. The 13 stars successfully observed and analyzed in this work are listed in Table~\ref{tab:observation_and_photometry_details}, together with their initial photometric metallicity estimates. At the time of target selection, information from previous high-resolution spectroscopic studies was not systematically compiled and was not used to select or prioritize targets. A subsequent literature review conducted during the preparation of this manuscript identified published high-resolution abundance analyses for a subset of our targets; comparisons with these studies are presented where relevant in the subsequent analysis and discussion.

High-resolution spectroscopic observations were carried out with the CFHT/ESPaDOnS spectrograph on August 28, September 1, November 23, and November 28, 2021. All targets were observed in the star+sky mode, providing a spectral resolving power of $\mathrm{R} \approx 68{,}000$ and a wavelength coverage of 3700--10\,500\,\AA. The resulting spectra reach a typical signal-to-noise ratio of $\mathrm{S/N} \sim 60$ at 5500\,\AA. The observational details for individual targets are presented in Table~\ref{tab:observation_and_photometry_details}. All spectra were reduced using the standard CFHT/ESPaDOnS reduction pipeline Upena\,\footnote{\url{https://www.cfht.hawaii.edu/Instruments/Upena/}} which is based on the Libre-ESpRIT package \citep{1997MNRAS.291..658D}. The reduction procedure includes bias subtraction, flat-field correction, order tracing, optimal spectral extraction, wavelength calibration, variance-weighted stitching of the spectral orders, barycentric correction, and subtraction of the sky spectrum in the star+sky observing mode.


Radial velocities (RVs) are measured from the reduced one-dimensional spectra using the cross-correlation function (CCF) method, and the derived values are listed in Table~\ref{tab:orbit_short}. Owing to the relatively low signal-to-noise ratio at the blue end of the spectra, we degrade the spectral resolution from $\mathrm{R} \approx 68{,}000$ to $\mathrm{R} \approx 47{,}000$, achieving a minimum $\mathrm{S/N} \sim 30$ around 4500\,\AA. Because several key neutron-capture lines, including Sr\,{\sc ii} and Eu\,{\sc ii} features, are located in the blue spectral region, we adopt a conservative approach for their abundance measurements: abundances are reported only when the relevant lines can be reliably identified and fitted, while no abundance values are given for stars with bad spectral quality or severe blending. After radial-velocity correction and spectral resolution degradation, each spectrum is locally normalized using a manually defined continuum to prepare the data for subsequent spectroscopic analysis.


\section{Spectral Analysis of Program Stars} \label{Spectrum Analysis of Program Stars}

The abundance analysis is performed with a homogeneous and largely automated pipeline implemented within the {\sc i}S{\sc pec} framework \citep{2014A&A...569A.111B,2019MNRAS.486.2075B}. Radiative-transfer calculations are carried out with the {\sc moog} code \citep{2012ascl.soft02009S} together with MARCS model atmospheres \citep{2008A&A...486..951G}. Individual abundances are derived either from equivalent-width measurements obtained through Gaussian profile fitting or from spectral synthesis calculations. We adopt the solar abundance scale of \citet{2007SSRv..130..105G}. Unless otherwise specified, all analyses and discussions in this work are based on the assumption of local thermodynamic equilibrium (LTE).

Stellar atmospheric parameters are determined through an iterative procedure designed to achieve internal consistency between the adopted model atmosphere and the observed spectral features. In this procedure, EW measurements of Fe lines are used to constrain the spectroscopic metallicity and check excitation and line-strength trends. With the final set of atmospheric parameters, elemental abundances are derived either from EW measurements or from spectral synthesis, depending on the availability and characteristics of the spectral lines. The details of these procedures are described in the following subsections.

\subsection{Stellar Atmospheric Parameters} \label{Stellar Atmospheric Parameters}


The effective temperature, $T_{\rm eff}$, is derived using the $(V-K_{\rm s})$ color--temperature relation from Table~3 of \citet{2005ApJ...626..465R}, which was calibrated with the infrared flux method (IRFM). Since several sample stars have relatively large uncertainties in their APASS $V$-band magnitudes \citep{2018AAS...23222306H}, we do not directly adopt the APASS photometry for the temperature determination. Instead, we derive homogeneous $(V-K_{\rm s})$ colors for program stars from SkyMapper DR2 $g$ photometry \citep{2019PASA...36...33O} and 2MASS $K_{\rm s}$ magnitudes \citep{2006AJ....131.1163S}, using an empirical $(g-K_{\rm s})$--$(V-K_{\rm s})$ transformation. The transformation is calibrated with standard stars from \citet{2013AJ....146...88C}, and the detailed fitting procedure is described in Appendix~\ref{color transformation}.

Reddening values, $E(B-V)$, are adopted from the two-dimensional dust map of \citet[][the SFD map]{1998ApJ...500..525S}, with the recalibration of \citet{2011ApJ...737..103S}. The extinction values adopted in this work are listed in Table~\ref{tab:observation_and_photometry_details}, together with the Galactic coordinates of the sample stars. Since all targets are located at relatively high Galactic latitudes ($|b| \geq 15^\circ$), the difference between two-dimensional and three-dimensional extinction estimates is expected to be small for the purpose of the present analysis. The extinction coefficients of $R_g = 3.407$ \citep{2019ApJS..243....7H} and $R_{K_{\rm s}} = 0.329$ \citep{2004AJ....128.2144M} are adopted. The resulting uncertainties in the derived effective temperatures are estimated to be approximately 35–50 K for all sample stars.

\begin{figure}
\centering
\includegraphics[width=1.0\linewidth]{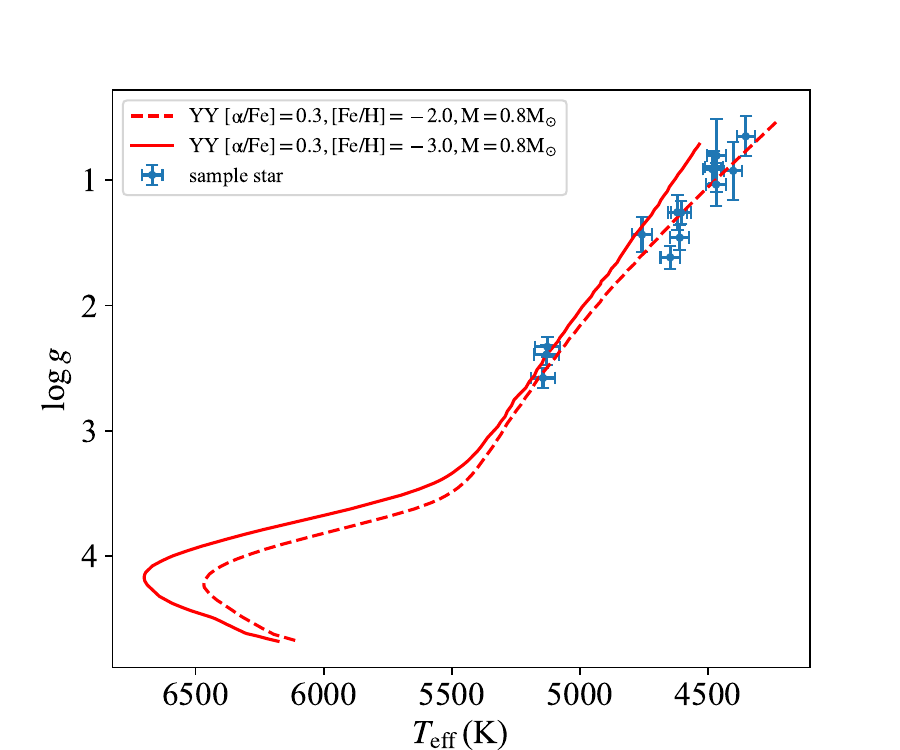}
\caption{The distribution of our sample stars in the $T_\mathrm{eff}$ vs.\ $\log g$ diagram. The 0.8\,$M_{\odot}$, alpha-enhanced Y$^2$ evolutionary tracks with [Fe/H]$=-2.0$ and [Fe/H]$=-3.0$ are shown as the red dashed and red solid lines, respectively. The sample stars are shown as blue dots.}
\label{fig:track}
\end{figure}

\begin{deluxetable}{ccccc}
\tabcolsep=0.35cm
\tabletypesize{\footnotesize}
\tablecaption{Determined Stellar Atmospheric Parameters} 
\label{Table of Stellar Atmospheric Parameters}
\tablehead{
\colhead{ID} & \colhead{$T_{\rm eff}$} & \colhead{log\,$g$} & \colhead{[Fe/H]} & \colhead{$v_{\rm mic}$} \\
\colhead{} & \colhead{(K)} & \colhead{} & \colhead{} & \colhead{(km\,s$^{-1}$)}
}
\startdata
J0858-0809 & 4606 & 1.26 & $-3.00$ & 1.75 \\
J1930-1739 & 4612 & 1.46 & $-2.59$ & 1.73 \\
J1944-0837 & 4649 & 1.62 & $-3.05$ & 1.51 \\
J1950-1255 & 4468 & 0.80 & $-2.88$ & 1.77 \\
J1952-1836 & 4403 & 0.93 & $-2.59$ & 1.73 \\
J2001-1215 & 4485 & 0.91 & $-2.52$ & 1.63 \\
J2020-1002 & 4470 & 1.03 & $-2.67$ & 1.64 \\
J2023-0815 & 5131 & 2.39 & $-2.75$ & 1.43 \\
J2026-0047 & 5128 & 2.33 & $-2.41$ & 1.25 \\
J2038-0023 & 4621 & 1.26 & $-2.97$ & 1.98 \\
J2043-1305 & 5145 & 2.58 & $-2.56$ & 1.26 \\
J2140-1227 & 4759 & 1.43 & $-3.00$ & 1.68 \\
J2326-0159 & 4477 & 0.90 & $-3.02$ & 1.91 \\
\enddata
\end{deluxetable}


The surface gravity, $\log g$, is calculated using the standard fundamental relation derived from the Stefan–Boltzmann law:
\begin{align}
\log g &= \log g_{\odot} + \log\left(\frac{M}{M_{\odot}}\right)
          + 4\,\log\left(\frac{T_{\rm eff}}{T_{\rm eff\odot}}\right) \notag \\
       &\quad + 0.4\,(M_{\rm bol}-M_{\rm bol\odot}) ,
\end{align}
where $M_{\rm bol}$ is the bolometric magnitude. $M_{\rm bol}$ is derived from the dereddened {\it Gaia} G-band magnitudes using $R_G=2.516$ \citep{2021ApJ...907...68H}, with the bolometric corrections from \citet{2018MNRAS.479L.102C} and the geometric distances from \citet{2021AJ....161..147B}.  The adopted distances are well constrained because all sample stars have {\it Gaia} parallax uncertainties below 20\% and lie within 7~kpc. Given the evolutionary stage of our sample stars, which are low-mass metal-poor giants, we adopt a fixed stellar mass of $M = 0.8\,M_{\odot}$ for all stars when computing surface gravities. Varying the stellar mass within a reasonable range for such stars leads to changes in $\log g$ of no more than $\sim$0.05\,dex, which is smaller than the typical uncertainty in the derived surface gravities. The solar reference values are $\log g_{\odot} = 4.44$, $T_{\rm eff\odot} = 5780$~K, and $M_{\rm bol\odot} = 4.74$. The uncertainty in $\log g$, propagated from all input parameters in the above equation, is smaller than 0.3~dex.

Since only a small number of Fe\,{\sc ii} lines (typically 2--5) are detectable in our spectra, we adopt the mean abundance derived from Fe\,{\sc i} lines as the stellar metallicity. Individual Fe\,{\sc i} abundances are computed from the measured equivalent widths through radiative-transfer calculations with {\sc moog} and the adopted MARCS model atmospheres, following the analysis framework described at the beginning of Section\,\ref{Spectrum Analysis of Program Stars}. The microturbulent velocity, $v_{\rm mic}$, is determined by adjusting its value until the abundances derived from Fe\,{\sc i} lines spanning a wide range of reduced equivalent widths show no systematic trend with line strength.


Because the metallicity affects the determination of $T_{\rm eff}$ and, in turn, the surface gravity $\log g$, we iterate the whole analysis to obtain a self-consistent set of stellar atmospheric parameters. In the first iteration, we adopt the photometric metallicity listed in Table~\ref{tab:observation_and_photometry_details} and an initial estimate of the microturbulent velocity from Equation~2 of \citet{2017Mashonkina_b}, together with the effective temperatures and surface gravities derived as described above. In subsequent iterations, the metallicity and microturbulent velocity obtained from the previous iteration are used as inputs, and the stellar parameters that depend on the adopted metallicity are updated accordingly. The $\alpha$-element abundance is fixed at [$\alpha$/Fe]$=+0.3$ throughout the iterations. This procedure is repeated until the metallicity difference between two successive iterations falls below 0.01\,dex. The full analysis is performed within the {\sc i}S{\sc pec} framework \citep{2014A&A...569A.111B, 2019MNRAS.486.2075B}. The final atmospheric parameters are listed in Table~\ref{Table of Stellar Atmospheric Parameters}. To compare the observed stellar parameters with theoretical predictions, we adopt Yonsei--Yale (Y$^2$) evolutionary tracks \citep{2001ApJS..136..417Y,2003ApJS..144..259Y}, interpolated to [Fe/H]$=-2.0$ and $-3.0$ for a stellar mass of $0.8\,M_\odot$ and [$\alpha$/Fe]$=+0.3$. As shown in Figure~\ref{fig:track}, the observed locations of the sample stars are in good agreement with the Y$^2$ model predictions.

Considering both the propagated uncertainties and potential systematic effects of the adopted methods, we adopt conservative uncertainties of 100\,K in $T_{\rm eff}$, 0.3\,dex in $\log g$, and 0.2\,km\,s$^{-1}$ in $v_{\rm mic}$. For the metallicity, the line-to-line scatter (standard deviation) of the Fe\,{\sc i}-based abundances is adopted as the uncertainty.


\subsection{Element Abundances} \label{Element Abundances}

\subsubsection{Atomic Parameters}


Reliable atomic parameters are fundamental to the determination of accurate elemental abundances. In this work, we have compiled a comprehensive set of atomic data for all targeted absorption lines, drawing from various literature sources as summarized in Table~\ref{linelist reference}. For Li and C, atomic parameters are adopted from the VALD database \citep{2015PhyS...90e4005R}. For most other species, the atomic data are taken from dedicated studies that focus on individual elements. For elements whose hyperfine structure is significant, including Mn, Sr, Ba, and Eu, we generate detailed line lists using the {\texttt{linemake}} code \citep{2021RNAAS...5...92P}. 


In the spectral-synthesis analysis, isotopic effects are treated according to the adopted line lists and the sensitivity of the relevant spectral features. For the CH molecular features used to derive carbon abundances, we directly adopt the molecular line data from VALD and do not attempt an independent determination of the $^{12}{\rm C}/^{13}{\rm C}$ ratio. For our sample stars, the CH features are relatively weak, and the typical signal-to-noise ratios in the CH absorption region are only $\sim$20--30, which prevents a reliable constraint on the carbon isotopic ratio. For barium, isotopic splitting is explicitly included, because the Ba\,{\sc ii}\,$\lambda$4554\,\AA\ resonance line used in this work is sensitive to the adopted isotopic composition \citep{2019AstL...45..341M}. We adopt an odd-isotope fraction of $f_{\mathrm{odd},r}=0.46$, representative of an \emph{r}-process-dominated isotopic mixture \citep{1999ApJ...521..691T,2006A&A...456..313M}.


\begin{deluxetable}{cc}
\tabcolsep=0.7cm
\tablecaption{Atomic Data References} \label{linelist reference}
\tablehead{Element & Reference}
\startdata
Li\,{\sc i} & \cite{2015PhyS...90e4005R} \\
CH  & \cite{2015PhyS...90e4005R} \\
Na\,{\sc i} & \cite{2024MNRAS.530.4712A} \\
Mg\,{\sc i} & \cite{2013Mashonkina} \\
Si\,{\sc i} & \cite{2016ApJ...833..225Z} \\
K\,{\sc i}  & \cite{2006Zhang} \\
Ca\,{\sc i} & \cite{2017Mashonkina_a} \\
Sc\,{\sc ii} & \cite{2008Zhang} \\
Ti\,{\sc i} \& Ti\,{\sc ii} & \cite{2017Mashonkina_b} \\
V\,{\sc i}  & \cite{2022ApJ...931..147L} \\
Cr\,{\sc i} & \cite{2010Bergemann} \\
Mn\,{\sc i} & \cite{2021RNAAS...5...92P} \\
Fe\,{\sc i} \& Fe\,{\sc ii} & \cite{2017Mashonkina_b} \\
Co\,{\sc i} & \cite{2010MNRAS.401.1334B} \\
Ni\,{\sc i} & \cite{2022ApJ...931..147L} \\
Zn\,{\sc i} & \cite{2022MNRAS.515.1510S} \\
Sr\,{\sc ii} & \cite{2021RNAAS...5...92P} \\
Y\,{\sc ii}  & \cite{2023ApJ...957...10A} \\
Zr\,{\sc ii} & \cite{2016ApJ...833..225Z} \\
Ba\,{\sc ii} & \cite{2021RNAAS...5...92P} \\
Eu\,{\sc ii} & \cite{2021RNAAS...5...92P} \\
\enddata
\end{deluxetable}

\par\vspace{-1.5\baselineskip}

\subsubsection{Abundance Determination}

Elemental abundances are derived using two complementary approaches: EW measurements and spectral synthesis. EWs are measured for absorption lines of 14 elements (Na, Mg, Si, K, Ca, Sc, Ti, V, Cr, Co, Ni, Zn, Y, and Zr), and line-by-line abundances are derived from these EWs using the same {\sc moog}/MARCS analysis framework described at the beginning of Section\,\ref{Spectrum Analysis of Program Stars}. For elements with multiple measurable transitions, we select lines with EWs of 20--100\,m\AA, avoiding both very weak lines, for which random measurement uncertainties become significant, and strong lines, which are more affected by saturation and microturbulence. When multiple lines are available for a given species, we adopt the mean of the individual line abundances as the final abundance. For species measured from a single usable line, the single-line abundance is adopted after verification by comparison with synthetic spectra.

\begin{figure*}[!htbp]
\centering
\includegraphics[scale=0.65]{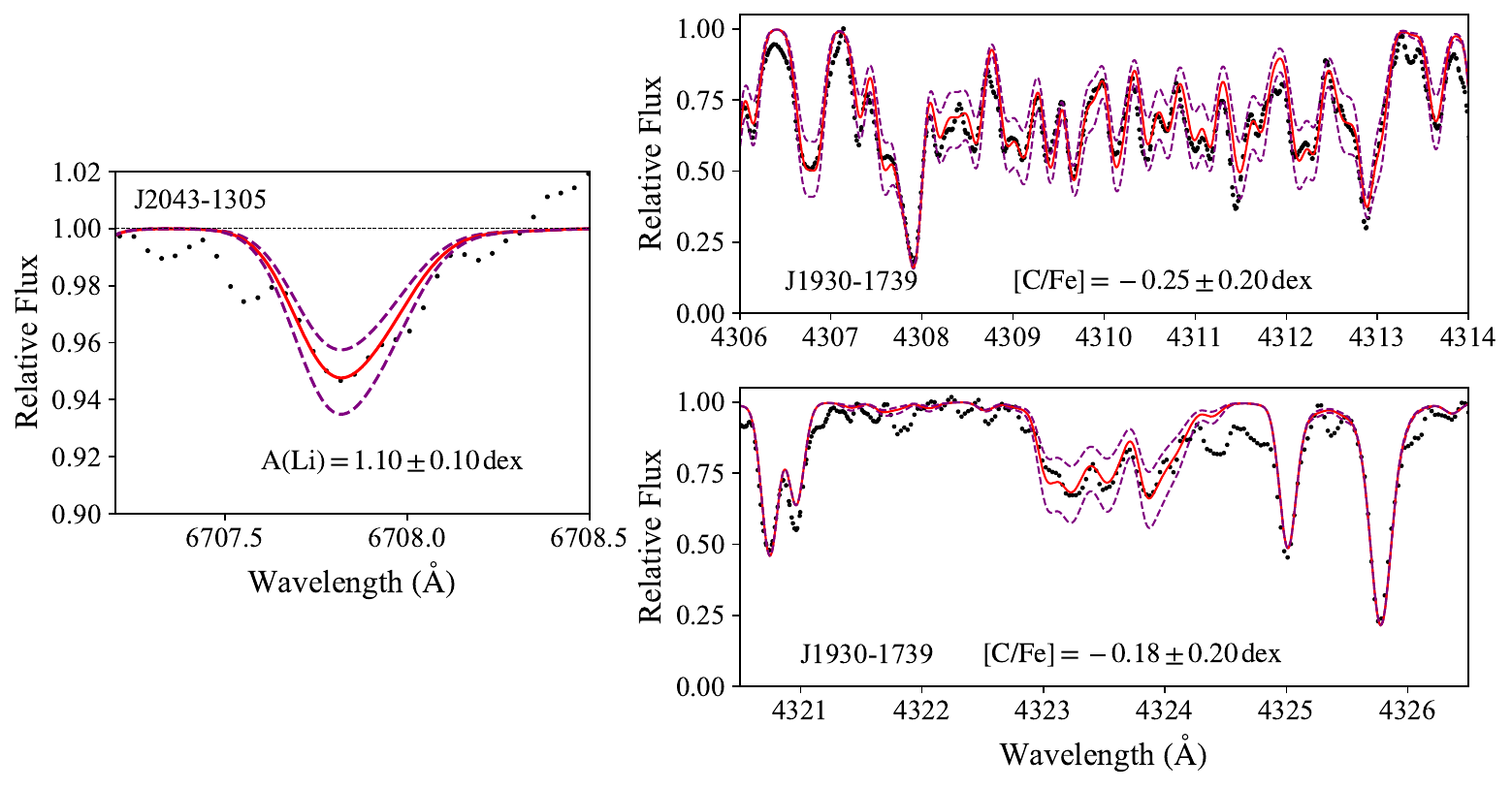}
\caption{{\it Left panel}: Spectrum fitting for the Li\,{\sc i}\,$\lambda$6707.8\,\AA\ line, using J2043-1305 as an example. The observed spectrum is shown as dots, while the best-fit profile and its 1$\sigma$ measurement uncertainty are indicated by the solid and dashed lines, respectively. {\it Right panel}: Spectrum fitting for the CH $G$-band, using J1930-1739 as an example. Two regions sensitive to the carbon abundance are shown. The symbols of the observed spectrum, the best-fit profile and its 1$\sigma$ measurement uncertainty are the same as left panel.}
\label{fig: synthesis fitting}
\end{figure*}

Spectral synthesis is adopted to derive abundances for Li\,{\sc i}\,$\lambda$6707.8\,\AA, carbon from the CH $G$-band region, and elements affected by hyperfine structure, including Mn, Sr, Ba, and Eu. Similar to above analysis, when multiple usable absorption lines or spectral regions are available, the final abundance is adopted as the average of the individual measurements; otherwise, the abundance from the single usable feature is adopted. In particular, the carbon abundance is independently measured from two CH spectral windows, 4306--4314\,\AA\ and 4321--4326\,\AA, and the final value is taken as their average. Examples of the spectral fitting for lithium and carbon are shown in Figure~\ref{fig: synthesis fitting}.

Because all stars in our sample are evolved giants, photospheric lithium and carbon abundances are reduced by the first dredge-up. As a result, these features are weak in most stars, and measurable lithium abundances are obtained for only three objects. To recover the natal carbon abundances, we apply evolutionary corrections following \citet{2014ApJ...797...21P}. Both the measured and corrected carbon abundances are listed in Table~\ref{tab: abundance}.

Uncertainties in the derived abundances arise from two main sources: measurement uncertainties and uncertainties propagated from the stellar atmospheric parameters. For lithium and carbon, measurement uncertainties are estimated by adopting upper and lower abundance limits from the best-fitting synthetic spectra. For elements with multiple measurable lines, the line-to-line scatter, calculated as the standard deviation of the abundances derived from individual lines, is adopted as the measurement uncertainty. For elements measured from a single line, a species specific line-to-line scatter cannot be determined. Given the large number of Fe\,{\sc i} lines measured across the spectrum, their line-to-line scatter provides a representative empirical measure of the overall spectral quality and the typical precision of individual line abundance measurements for each star. We therefore adopt the Fe\,{\sc i} line-to-line scatter as the measurement uncertainty for such elements, following similar treatments in previous high-resolution abundance analyses of metal-poor stars \citep[e.g.,][]{2010ApJ...711..350N,2022ApJ...931..147L}. All measurement uncertainties are reported in Table~\ref{tab: abundance}.

\begin{longrotatetable}
\begin{deluxetable*}{lcccc cccc cccc cccc cccc cccc cccc c} \tablecaption{Abundances and Abundance Errors for Sample Stars \label{tab: abundance}} 
\tabletypesize{\scriptsize} 
\tablehead{\vspace{-6.5ex}}  
\makeatother
\startdata 
& \multicolumn{3}{c}{\raisebox{-1ex}{J0858-0809}} & & \multicolumn{3}{c}{\raisebox{-1ex}{J1930-1739}} & & \multicolumn{3}{c}{\raisebox{-1ex}{J1944-0837}} & & \multicolumn{3}{c}{\raisebox{-1ex}{J1950-1255}} & & \multicolumn{3}{c}{\raisebox{-1ex}{J1952-1836}} & &
\multicolumn{3}{c}{\raisebox{-1ex}{J2001-1215}} & &
\multicolumn{3}{c}{\raisebox{-1ex}{J2020-1002}} & \\ [6pt]
\cline{2-4} \cline{6-8} \cline{10-12} \cline{14-16} \cline{18-20} \cline{22-24} \cline{26-28} \\[-6pt] 
\colhead{Ion} & \colhead{[X/Fe]$^{a}$} & \colhead{$\sigma$(N)$^{b}$} & \colhead{N$^{c}$} & & \colhead{[X/Fe]} & \colhead{$\sigma$(N)} & \colhead{N} & & \colhead{[X/Fe]} & \colhead{$\sigma$(N)} & \colhead{N} & & \colhead{[X/Fe]} & \colhead{$\sigma$(N)} & \colhead{N} & & \colhead{[X/Fe]} & \colhead{$\sigma$(N)} & \colhead{N} & & \colhead{[X/Fe]} & \colhead{$\sigma$(N)} & \colhead{N} & & \colhead{[X/Fe]} & \colhead{$\sigma$(N)} & \colhead{N} & \\
\hline
A(Li) & \nodata & \nodata & \nodata & &\nodata & \nodata & \nodata & &\nodata & \nodata & \nodata & &\nodata & \nodata & \nodata & &\nodata & \nodata & \nodata & &\nodata & \nodata & \nodata & &\nodata & \nodata & \nodata \\
CH & $-$0.05 & 0.30 & 2 & &$-$0.22 & 0.20 & 2 & &$-$0.05 & 0.30 & 2 & &$-$0.30 & 0.30 & 1 & &$-$0.75 & 0.30 & 2 & &$-$0.36 & 0.30 & 2 & &$-$0.40 & 0.40 & 2 \\
$\mathrm{CH_{cor}}$ & 0.54 & \nodata & \nodata & &0.25 & \nodata & \nodata & &0.28 & \nodata & \nodata & & 0.44 & \nodata & \nodata & &0.03 & \nodata & \nodata & &0.38 & \nodata & \nodata & &0.34 & \nodata & \nodata \\
\ion{Na}{1} & 0.35 & 0.04 & 2 & &0.35 & 0.01 & 2 & &0.33 & 0.07 & 2 & &0.24 & 0.02 & 2 & &0.26 & 0.03 & 2 & &\nodata & \nodata & \nodata & &0.24 & 0.04 & 2 \\
\ion{Mg}{1} & 0.57 & 0.12 & 5 & &0.42 & 0.15 & 4 & &0.47 & 0.05 & 5 & &0.52 & 0.03 & 5 & &0.44 & 0.02 & 3 & &0.53 & 0.11 & 5 & &0.51 & 0.11 & 5 \\
\ion{Si}{1} & \nodata & \nodata & \nodata & &\nodata & \nodata & \nodata & &\nodata & \nodata & \nodata & &\nodata & \nodata & \nodata & &0.50 & 0.09 & 1 & &\nodata & \nodata & \nodata & &\nodata & \nodata & \nodata \\
\ion{K}{1} & 0.55 & 0.13 & 1 & &0.50 & 0.02 & 2 & &0.37 & 0.11 & 1 & &0.31 & 0.12 & 1 & &0.47 & 0.09 & 1 & &1.06$^{d}$ & 0.10 & 1 & &0.48 & 0.12 & 1 \\
\ion{Ca}{1} & 0.33 & 0.06 & 6 & &0.32 & 0.09 & 9 & &0.35 & 0.08 & 4 & &0.30 & 0.14 & 6 & &0.37 & 0.06 & 10 & &0.32 & 0.09 & 11 & &0.34 & 0.11 & 8 \\
\ion{Sc}{2} & 0.13 & 0.01 & 2 & &0.17 & 0.11 & 5 & &0.22 & 0.20 & 4 & &0.06 & 0.15 & 5 & &0.22 & 0.09 & 7 & &0.09 & 0.12 & 8 & &0.11 & 0.15 & 4 \\
\ion{Ti}{1} & 0.23 & 0.07 & 9 & &0.17 & 0.07 & 11 & &0.18 & 0.03 & 8 & &0.08 & 0.07 & 10 & &0.02 & 0.09 & 11 & &0.11 & 0.08 & 12 & &0.09 & 0.07 & 11 \\
\ion{Ti}{2} & 0.51 & 0.12 & 8 & &0.55 & 0.11 & 10 & &0.54 & 0.07 & 6 & &0.43 & 0.12 & 8 & &0.69 & 0.13 & 8 & &0.45 & 0.09 & 7 & &0.52 & 0.06 & 6 \\
\ion{V}{1} & 0.03 & 0.13 & 1 & &$-$0.22 & 0.10 & 1 & &$-$0.05 & 0.11 & 1 & &$-$0.30 & 0.12 & 1 & &$-$0.33 & 0.09 & 1 & &$-$0.13 & 0.10 & 1 & &$-$0.07 & 0.12 & 1 \\
\ion{Cr}{1} & $-$0.11 & 0.13 & 1 & &$-$0.25 & 0.12 & 6 & &\nodata & \nodata & \nodata & &$-$0.25 & 0.06 & 4 & &$-$0.31 & 0.14 & 9 & &$-$0.27 & 0.10 & 7 & &$-$0.34 & 0.11 & 6 \\
\ion{Mn}{1} & \nodata & \nodata & \nodata & &$-$0.39 & 0.09 & 3 & &\nodata & \nodata & \nodata & &$-$0.37 & 0.07 & 3 & &$-$0.54 & 0.10 & 3 & &$-$0.43 & 0.04 & 2 & &$-$0.59 & 0.09 & 2 \\
\ion{Fe}{1} & 0.00 & 0.13 & 42 & & 0.00 & 0.10 & 51 & & 0.00 & 0.11 & 37 & & 0.00 & 0.12 & 46 & & 0.00 & 0.09 & 51 & & 0.00 & 0.10 & 53 & & 0.00 & 0.12 & 44 \\
\ion{Fe}{2} & 0.23 & 0.05 & 4 & & 0.23 & 0.04 & 4 & & 0.42 & 0.10 & 4 & & 0.17 & 0.03 & 2 & & 0.30 & 0.12 & 5 & & 0.19 & 0.10 & 5 & & 0.21 & 0.02 & 2 \\
\ion{Co}{1} & \nodata & \nodata & \nodata & &0.25 & 0.10 & 1 & &\nodata & \nodata & \nodata & &\nodata & \nodata & \nodata & &\nodata & \nodata & \nodata & &\nodata & \nodata & \nodata & &0.43 & 0.02 & 2 \\
\ion{Ni}{1} & 0.03 & 0.22 & 2 & &0.03 & 0.12 & 9 & &$-$0.15 & 0.11 & 1 & &$-$0.03 & 0.08 & 4 & &$-$0.01 & 0.09 & 8 & &0.02 & 0.15 & 9 & &$-$0.06 & 0.08 & 4 \\
\ion{Zn}{1} & \nodata & \nodata & \nodata & &0.46 & 0.10 & 1 & &\nodata & \nodata & \nodata & &0.47 & 0.07 & 2 & &0.33 & 0.07 & 2 & &0.20 & 0.10 & 1 & &0.48 & 0.12 & 1 \\
\ion{Sr}{2} & \nodata & \nodata & \nodata & &0.62 & 0.11 & 2 & &\nodata & \nodata & \nodata & &0.32 & 0.12 & 1 & &0.75 & 0.09 & 1 & &0.40 & 0.10 & 1 & &0.32 & 0.12 & 1 \\
\ion{Y}{2} & \nodata & \nodata & \nodata & &0.09 & 0.06 & 4 & &0.28 & 0.18 & 3 & &$-$0.22 & 0.01 & 2 & &0.06 & 0.11 & 5 & &$-$0.18 & 0.05 & 4 & &$-$0.02 & 0.15 & 2 \\
\ion{Zr}{2} & \nodata & \nodata & \nodata & &0.53 & 0.04 & 2 & &\nodata & \nodata & \nodata & &\nodata & \nodata & \nodata & &\nodata & \nodata & \nodata & &\nodata & \nodata & \nodata & &\nodata & \nodata & \nodata \\
\ion{Ba}{2} & $-$0.13 & 0.02 & 2 & &$-$0.02 & 0.02 & 2 & &$-$0.56 & 0.05 & 2 & &$-$0.60 & 0.04 & 2 & &$-$0.05 & 0.17 & 2 & &0.02 & 0.12 & 2 & &0.23 & 0.07 & 2 \\
\ion{Eu}{2} & 0.57 & 0.13 & 1 & &0.50 & 0.10 & 1 & &\nodata & \nodata & \nodata & &\nodata & \nodata & \nodata & &\nodata & \nodata & \nodata & &0.87 & 0.10 & 1 & &0.43 & 0.12 & 1 \\
\hline
\hline
& \multicolumn{3}{c}{\raisebox{-1ex}{J2023-0815}} & & \multicolumn{3}{c}{\raisebox{-1ex}{J2026-0047}} & & \multicolumn{3}{c}{\raisebox{-1ex}{J2038-0023}} & & \multicolumn{3}{c}{\raisebox{-1ex}{J2043-1305}} & & \multicolumn{3}{c}{\raisebox{-1ex}{J2140-1227}} & &
\multicolumn{3}{c}{\raisebox{-1ex}{J2326-0159}} & \\ [6pt]
\cline{2-4} \cline{6-8} \cline{10-12} \cline{14-16} \cline{18-20} \cline{22-24} \\[-6pt] 
\colhead{Ion} & \colhead{[X/Fe]} & \colhead{$\sigma$(N)} & \colhead{N} & & \colhead{[X/Fe]} & \colhead{$\sigma$(N)} & \colhead{N} & & \colhead{[X/Fe]} & \colhead{$\sigma$(N)} & \colhead{N} & & \colhead{[X/Fe]} & \colhead{$\sigma$(N)} & \colhead{N} & & \colhead{[X/Fe]} & \colhead{$\sigma$(N)} & \colhead{N} & & \colhead{[X/Fe]} & \colhead{$\sigma$(N)} & \colhead{N} & \\
\hline
A(Li) & 1.06 & 0.10 & 1 & &0.90 & 0.20 & 1 & &\nodata & \nodata & \nodata & &1.10 & 0.10 & 1 & &\nodata & \nodata & \nodata & &\nodata & \nodata & \nodata \\
CH & 0.35 & 0.30 & 2 & &0.24 & 0.30 & 2 & &$-$0.32 & 0.30 & 1 & &0.43 & 0.30 & 2 & &0.38 & 0.20 & 2 & &$-$0.30 & 0.30 & 1 \\
$\mathrm{CH_{cor}}$ & 0.36 & \nodata & \nodata & & 0.25 & \nodata & \nodata & & 0.28 & \nodata & \nodata & &0.44 & \nodata & \nodata & &0.85 & \nodata & \nodata & &0.44 & \nodata & \nodata \\
\ion{Na}{1} & 0.32 & 0.03 & 2 & &0.32 & 0.04 & 2 & &0.30 & 0.11 & 1 & &0.27 & 0.05 & 2 & &0.87 & 0.02 & 2 & &0.38 & 0.09 & 2 \\
\ion{Mg}{1} & 0.30 & 0.05 & 5 & &0.25 & 0.05 & 5 & &0.42 & 0.02 & 5 & &0.29 & 0.09 & 5 & &0.68 & 0.02 & 5 & &0.67 & 0.06 & 5 \\
\ion{Si}{1} & \nodata & \nodata & \nodata & &\nodata & \nodata & \nodata & &0.42 & 0.11 & 1 & &\nodata & \nodata & \nodata & &\nodata & \nodata & \nodata & &\nodata & \nodata & \nodata \\
\ion{K}{1} & \nodata & \nodata & \nodata & &0.49 & 0.13 & 1 & &0.42 & 0.01 & 2 & &0.54 & 0.12 & 1 & &0.72 & 0.14 & 2 & &0.45 & 0.10 & 1 \\
\ion{Ca}{1} & 0.26 & 0.10 & 5 & &0.25 & 0.09 & 7 & &0.24 & 0.10 & 6 & &0.25 & 0.06 & 6 & &0.33 & 0.05 & 4 & &0.35 & 0.08 & 5 \\
\ion{Sc}{2} & 0.20 & 0.11 & 3 & &0.30 & 0.09 & 2 & &0.10 & 0.13 & 5 & &0.13 & 0.12 & 3 & &0.03 & 0.05 & 3 & &0.15 & 0.12 & 5 \\
\ion{Ti}{1} & 0.21 & 0.07 & 4 & &0.22 & 0.07 & 6 & &0.14 & 0.10 & 7 & &0.24 & 0.09 & 7 & &0.38 & 0.06 & 9 & &0.22 & 0.07 & 9 \\
\ion{Ti}{2} & 0.42 & 0.02 & 3 & &0.39 & 0.12 & 5 & &0.45 & 0.14 & 5 & &0.38 & 0.05 & 4 & &0.61 & 0.10 & 9 & &0.53 & 0.10 & 9 \\
\ion{V}{1} & \nodata & \nodata & \nodata & &0.06 & 0.13 & 1 & &\nodata & \nodata & \nodata & &\nodata & \nodata & \nodata & &0.29 & 0.10 & 1 & &\nodata & \nodata & \nodata \\
\ion{Cr}{1} & \nodata & \nodata & \nodata & &$-$0.20 & 0.07 & 3 & &$-$0.27 & 0.04 & 2 & &$-$0.31 & 0.12 & 1 & &0.23 & 0.12 & 7 & &$-$0.25 & 0.04 & 2 \\
\ion{Mn}{1} & \nodata & \nodata & \nodata & &$-$0.33 & 0.13 & 1 & &\nodata & \nodata & \nodata & &\nodata & \nodata & \nodata & &0.00 & 0.01 & 2 & &\nodata & \nodata & \nodata \\
\ion{Fe}{1} & 0.00 & 0.09 & 33 & & 0.00 & 0.13 & 48 & & 0.00 & 0.11 & 37 & & 0.00 & 0.12 & 40 & & 0.00 & 0.10 & 35 & & 0.00 & 0.10 & 41 \\
\ion{Fe}{2} & 0.19 & 0.05 & 4 & & 0.21 & 0.12 & 3 & & 0.15 & 0.05 & 3 & & 0.25 & 0.08 & 4 & & 0.19 & 0.04 & 4 & & 0.22 & 0.12 & 3 \\
\ion{Co}{1} & 0.56 & 0.09 & 1 & &\nodata & \nodata & \nodata & &\nodata & \nodata & \nodata & &0.69 & 0.12 & 1 & &0.52 & 0.07 & 2 & &\nodata & \nodata & \nodata \\
\ion{Ni}{1} & 0.09 & 0.05 & 2 & &0.13 & 0.12 & 5 & &$-$0.13 & 0.11 & 1 & &$-$0.01 & 0.03 & 2 & &0.41 & 0.09 & 6 & &$-$0.13 & 0.10 & 1 \\
\ion{Zn}{1} & \nodata & \nodata & \nodata & &0.23 & 0.10 & 2 & &\nodata & \nodata & \nodata & &\nodata & \nodata & \nodata & &1.17 & 0.04 & 2 & &\nodata & \nodata & \nodata \\
\ion{Sr}{2} & $-$0.53 & 0.18 & 2 & &0.09 & 0.20 & 2 & &0.55 & 0.12 & 1 & &0.33 & 0.11 & 2 & & 1.69 & 0.10 & 1 & &0.55 & 0.10 & 1 \\
\ion{Y}{2} & \nodata & \nodata & \nodata & &0.20 & 0.03 & 2 & &0.35 & 0.09 & 4 & &\nodata & \nodata & \nodata & &0.82 & 0.09 & 7 & &0.15 & 0.14 & 5 \\
\ion{Zr}{2} & \nodata & \nodata & \nodata & &\nodata & \nodata & \nodata & &0.73 & 0.05 & 2 & &\nodata & \nodata & \nodata & &1.03 & 0.08 & 2 & &\nodata & \nodata & \nodata \\
\ion{Ba}{2} & $-$1.15 & 0.09 & 1 & &$-$0.06 & 0.10 & 2 & &0.76 & 0.06 & 2 & &$-$0.03 & 0.01 & 2 & &$-$0.28 & 0.05 & 3 & &0.23 & 0.07 & 2 \\
\ion{Eu}{2} & \nodata & \nodata & \nodata & &\nodata & \nodata & \nodata & &1.60 & 0.11 & 1 & &0.56 & 0.12 & 1 & &\nodata & \nodata & \nodata & &0.85 & 0.10 & 1 \\
\enddata 
\tablecomments{
$^{a}$ The [X/Fe] values represent the derived chemical abundances in unit of dex. Abundances of Na, Mg, Si, K, Ca, Sc, Ti, V, Cr, Co, Ni, Zn, Y, and Zr are derived from EW measurements, while those of Li, CH, Mn, Sr, Ba, and Eu are derived from spectral synthesis. For EW-based species with only a single usable line, the abundance is first derived from the EW measurement and then verified through comparison with synthetic spectra. \\
$^{b}$ The $\sigma$(N) values represent the statistical measurement uncertainties in unit of dex. For Li and C, the uncertainties correspond to the 1-$\sigma$ measurement uncertainties from spectral synthesis. For elements with more than one measurable line, they are calculated from the line-to-line scatter (standard deviation) of the derived abundances. For elements measured from a single line, the Fe\,{\sc i} line-to-line scatter is adopted as a representative measurement uncertainty. \\
$^{c}$ Number of spectral lines used in the abundance determination. \\
$^{d}$ For J2001-1215, NLTE analysis is applied to the potassium abundance. The adopted NLTE correction is $-0.29$\,dex, yielding $\mathrm{[K/Fe]_{NLTE}=0.77\pm0.10}$. The statistical uncertainty is adopted from the LTE abundance determination and does not include possible systematic uncertainties associated with the NLTE correction itself. \\}
The full table is also available in machine-readable form.
\end{deluxetable*}
\end{longrotatetable}


To evaluate the sensitivity of the derived abundances to uncertainties in the stellar atmospheric parameters, we recompute the abundances by varying each parameter independently by representative amounts: $\pm100$\,K in $T_{\rm eff}$, $\pm0.3$\,dex in $\log g$, and $\pm0.2$\,km\,s$^{-1}$ in $v_{\rm mic}$. The total abundance uncertainties are calculated by adding in quadrature the measurement uncertainties and the abundance variations caused by the adopted uncertainties in the stellar atmospheric parameters. The resulting abundance responses are summarized in Table~\ref{abundance uncertainties}. In general, the total abundance uncertainties for our sample stars are typically $\sim$0.15\,dex.

\begin{deluxetable}{lccccc}
\centering
\tabcolsep=0.25cm
\tablecaption{\centering{Example abundance uncertainties for J1930-1739}} \label{abundance uncertainties}
\tablehead{Ion & $\sigma$(N) & $\Delta T_\mathrm{eff}$ & $\Delta$log\,$g$ & $\Delta v_\mathrm{mic}$ & Total \\ 
  & (dex)  &  ($\pm100$\,K)    & ($\pm0.3$\,dex)   & ($\pm0.2$\,km\,s$^{-1}$) & (dex) }
\startdata
CH           & 0.20    & $\pm$0.17  & $\mp$0.10  & $\pm$0.00  & 0.28  \\
Na\,{\sc i}  & 0.01  & $\pm$0.19  & $\mp$0.10  & $\mp$0.10  & 0.24  \\
Mg\,{\sc i}  & 0.15  & $\pm$0.16  & $\mp$0.12  & $\mp$0.06  & 0.26  \\
K\,{\sc i}   & 0.02  & $\pm$0.11  & $\mp$0.02  & $\mp$0.04  & 0.12  \\
Ca\,{\sc i}  & 0.09  & $\pm$0.08  & $\mp$0.04  & $\mp$0.04  & 0.13  \\
Sc\,{\sc ii} & 0.11  & $\pm$0.01  & $\pm$0.09  & $\mp$0.05  & 0.15  \\
Ti\,{\sc i}  & 0.07  & $\pm$0.16  & $\mp$0.04  & $\mp$0.04  & 0.18  \\
Ti\,{\sc ii} & 0.11  & $\pm$0.01  & $\pm$0.08  & $\mp$0.04  & 0.14  \\
V\,{\sc i}   & 0.10  & $\pm$0.16  & $\mp$0.05  & $\mp$0.02  & 0.20  \\
Cr\,{\sc i}  & 0.12  & $\pm$0.14  & $\mp$0.04  & $\mp$0.02  & 0.19  \\
Mn\,{\sc i}  & 0.09  & $\pm$0.12  & $\mp$0.04  & $\pm$0.02  & 0.16  \\
Co\,{\sc i}  & 0.10  & $\pm$0.18  & $\mp$0.06  & $\mp$0.15  & 0.26  \\
Ni\,{\sc i}  & 0.12  & $\pm$0.11  & $\mp$0.02  & $\mp$0.03  & 0.17  \\
Zn\,{\sc i}  & 0.10  & $\pm$0.00  & $\pm$0.05  & $\mp$0.02  & 0.11  \\
Sr\,{\sc ii} & 0.11  & $\pm$0.10  & $\mp$0.11  & $\mp$0.08  & 0.20  \\
Y\,{\sc ii}  & 0.06  & $\pm$0.03  & $\pm$0.09  & $\mp$0.03  & 0.12  \\
Zr\,{\sc ii} & 0.04  & $\pm$0.02  & $\pm$0.07  & $\mp$0.05  & 0.10  \\
Ba\,{\sc ii} & 0.02  & $\pm$0.05  & $\pm$0.10  & $\mp$0.07  & 0.13  \\
Eu\,{\sc ii} & 0.10  & $\pm$0.05  & $\pm$0.08  & $\pm$0.00  & 0.14  \\
\enddata
\tablecomments{
The $\sigma$(N) values represent the statistical measurement uncertainties, following the same definitions as in Table\,\ref{tab: abundance}. The total uncertainties, including both statistical and systematic contributions propagated from the stellar atmospheric parameters, are presented in the last column. \\
The complete abundance uncertainty table for all 13 program stars is available in machine-readable format, which includes an additional ID column identifying each star.
}
\end{deluxetable}

\subsection{Validation of the Metallicity}

\subsubsection{Photometric versus High-resolution Spectroscopic Metallicities}

\begin{figure*}[!htbp]
\centering
\includegraphics[width=0.9\linewidth]{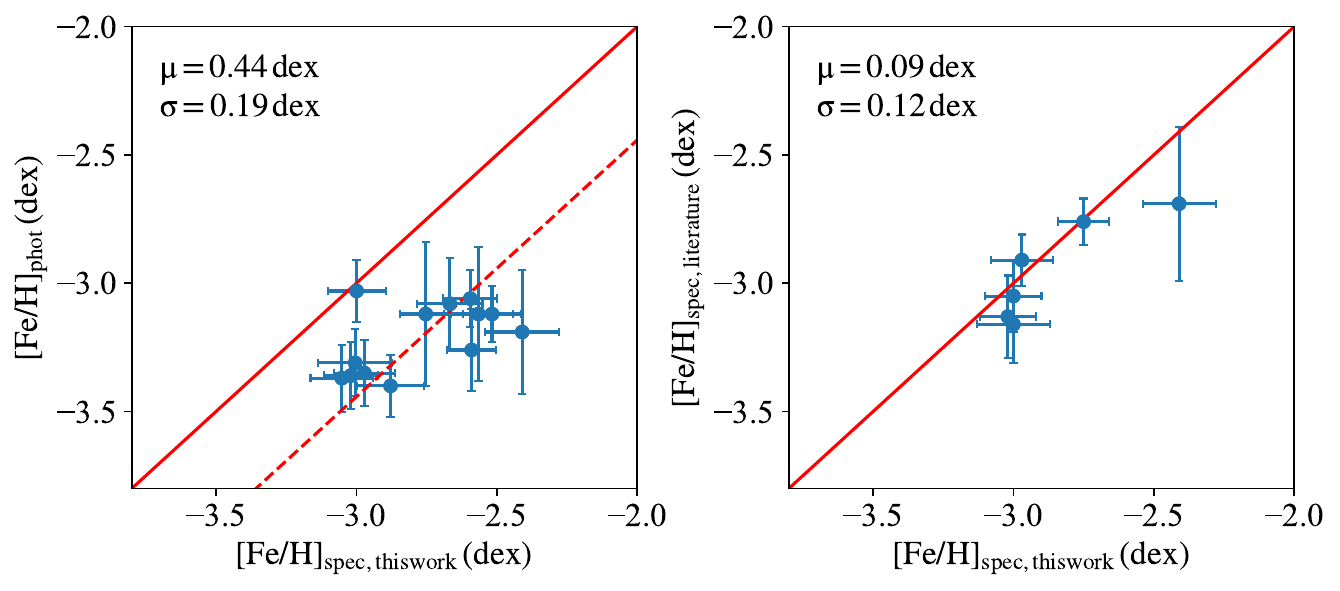}
\caption{
Comparison of metallicity estimates for the program stars.
Left: photometric metallicities used for the initial candidate selection compared with the high-resolution spectroscopic metallicities derived in this work.
Right: literature spectroscopic metallicities compared with those derived in this work for stars with previous high-resolution studies.
The solid red lines indicate one-to-one agreement.
In the left panel, the dashed red line shows the mean offset between the photometric and spectroscopic metallicities.
The mean offset and standard deviation are labeled in the top-left corner of each panel.
}
\label{fig: FEH_comparison}
\end{figure*}

Based on the atmospheric parameters and metallicities derived in Section\,\ref{Stellar Atmospheric Parameters}, we compare the photometric metallicities used for the initial candidate selection with the high-resolution spectroscopic metallicities derived in this work. As shown in Figure~\ref{fig: FEH_comparison}, the photometric metallicities systematically underestimate the spectroscopic values by 0.44\,dex, with a standard deviation of 0.19\,dex. The relatively small scatter indicates that the narrow-band photometric metallicities are effective for ranking metal-poor candidates, whereas the systematic offset suggests that further calibration is required for accurate EMP classification in the very metal-poor regime. Based on the spectroscopic metallicities, four out of the 13 candidates are confirmed as EMP stars with $\mathrm{[Fe/H]} < -3.0$, corresponding to an EMP confirmation rate of approximately 30\%, and all targets remain very metal-poor with $\mathrm{[Fe/H]} < -2.4$. If contamination is defined as stars with $\mathrm{[Fe/H]} > -2.0$, no such contaminants are present in the current sample. Despite the relatively small sample size, these results indicate that the narrow-band photometric selection efficiently concentrates candidates into the VMP regime while retaining a substantial EMP fraction.

\begin{table*}
\centering
\caption{Stellar atmospheric parameters for our sample stars in literature}
\label{tab:literature_parameter}
\begin{tabular}{lccccc}
\hline\hline
ID & source$^{a}$ & $T_{\rm eff}$ (K) & $\log g$ & [Fe/H] & $v_{\rm mic}$ (km\,s$^{-1}$) \\
\hline
J0858-0809 &   \citet{2018ApJ...864...43C}            & $4530 \pm 150$ & $0.70 \pm 0.30$ & $-3.16 \pm 0.15$ & $2.25 \pm 0.30$ \\
J2023-0815 & \citet{2020ApJ...898..150E}   & $4945 \pm 150$ & $1.63 \pm 0.30$ & $-2.76 \pm 0.09$ & $1.73 \pm 0.20$ \\
J2026-0047 & \citet{2021MNRAS.507.4102Y}   & $4825 \pm 100$ & $1.56 \pm 0.30$ & $-2.69 \pm 0.30$ & $1.30 \pm 0.30$ \\
J2038-0023 &  \citet{2017ApJ...844...18P}           & $4630 \pm 100$ & $1.20 \pm 0.20$ & $-2.91 \pm 0.10$ & $2.15 \pm 0.20$ \\
J2140-1227 & \citet{2024Dornbusch} & $4855 \pm 64$  & $1.44 \pm 0.12$ & $-3.05 \pm 0.14$ & $2.02 \pm 0.06$ \\
J2326-0159 & \citet{2018ApJ...858...92H}          & $4360 \pm 150$ & $0.45 \pm 0.30$ & $-3.13 \pm 0.16$ & $2.40 \pm 0.30$ \\
\hline
\end{tabular}
\vspace{0.2cm}
\tablecomments{
$^{a}$ For stars with multiple high-resolution studies available in the literature, we adopt the most recent LTE analysis to ensure a consistent comparison with the LTE abundance analysis presented in this work.
}
\end{table*}

\subsubsection{Comparison with Previous High-resolution Studies}

After cross-matching our sample with the SIMBAD database\footnote{\url{https://simbad.u-strasbg.fr/simbad/}}, we identify six stars that have previously been studied with high-resolution spectroscopy in the literature. As shown in Figure~\ref{fig: FEH_comparison}, the spectroscopic metallicities derived in this work are generally consistent with the literature values, with a mean offset of 0.09\,dex and a standard deviation of 0.12\,dex. This agreement indicates that our metallicity scale is broadly consistent with previous high-resolution analyses.

Although the metallicities show good agreement, the literature atmospheric parameters listed in Table~\ref{tab:literature_parameter} show somewhat larger scatter relative to those adopted in this work, particularly in $T_{\rm eff}$ and $\log g$ for some stars. These offsets largely arise from differences in the methods used to determine the stellar parameters. For J0858-0809, J2026-0047, and J2326-0159, the differences are primarily caused by the use of photometric temperatures and {\it Gaia} DR3 distance-based gravities in this work, compared with spectroscopic temperatures from excitation equilibrium and gravities from ionization balance in previous LTE-based studies. For J2023-0815, the difference is instead mainly driven by the adopted distance: \citet{2020ApJ...898..150E} used a {\it Gaia} DR2 distance exceeding 6\,kpc, whereas we adopt a {\it Gaia} DR3 geometric distance of 2.2\,kpc. In contrast, J2140-1227 was analyzed by \citet{2024Dornbusch} using photometric temperatures and {\it Gaia} DR3 distance-based gravities, closely matching our approach, and the resulting atmospheric parameters agree well with our values. We therefore regard the offsets in $T_{\rm eff}$ and $\log g$ primarily as methodological differences rather than evidence for inconsistencies in the abundance analysis; their possible effects on individual abundance comparisons are discussed further in Section~\ref{Discussion}.

\begin{figure*}[!htbp]
\centering
\includegraphics[width=0.9\linewidth]{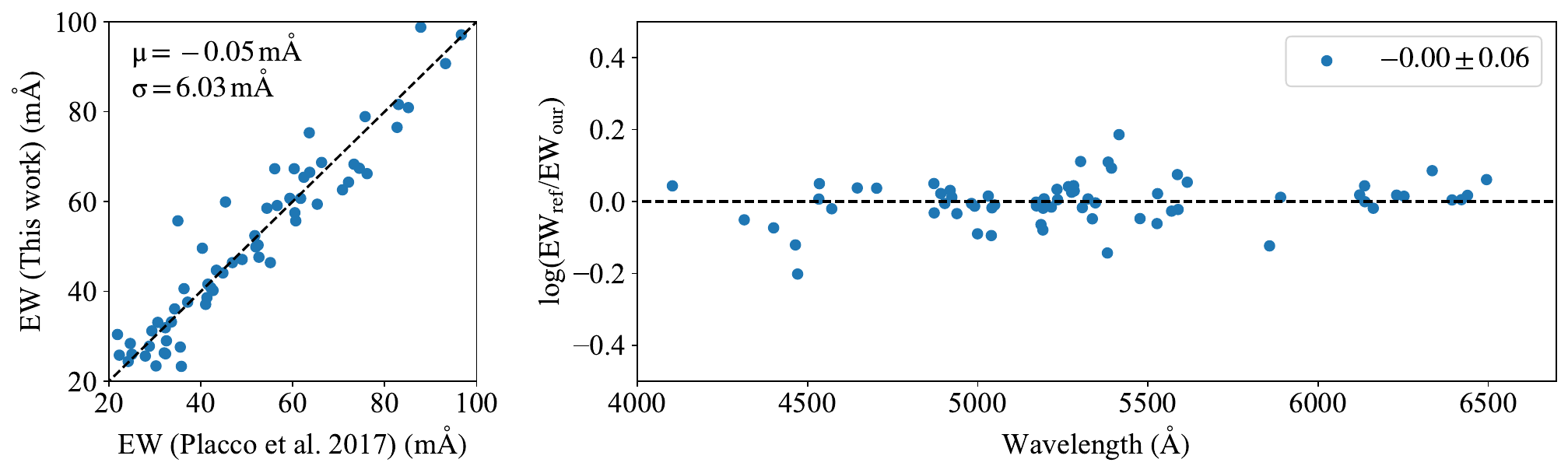}
\caption{Comparison of the equivalent widths (EWs) measured for J2038-0023 in this work with those reported by \citet{2017ApJ...844...18P}. Left panel: direct comparison of the measured EWs. The dashed black line indicates one-to-one agreement, and the mean offset and standard deviation are shown in the top-left corner. Right panel: logarithmic EW differences as a function of wavelength, defined as $\log(\mathrm{EW}_{\rm ref}/\mathrm{EW}_{\rm our})$, where $\mathrm{EW}_{\rm ref}$ denotes the values reported by \citet{2017ApJ...844...18P}. The mean offset and standard deviation are shown in the top-right corner.}
\label{fig:EW_comparison}
\end{figure*}

As an additional check on the abundance measurements themselves, we compare our equivalent-width (EW) measurements with literature values obtained from high-resolution spectra. We use J2038-0023 as a representative comparison object because the spectrum analyzed by \citet{2017ApJ...844...18P} has particularly high quality ($\mathrm{S/N}\sim100$ at 3860\,\AA\ and $\mathrm{S/N}\sim220$ at 4550\,\AA). As shown in Figure~\ref{fig:EW_comparison}, the EW measurements are in good agreement, with a mean offset of $-0.05$\,m\AA\ and a standard deviation of $6.03$\,m\AA. This agreement supports the reliability of our EW measurements.

\begin{longrotatetable}
\setlength{\tabcolsep}{2pt}
\begin{deluxetable*}{lrrrrrrrrrrrrrr}
\tablecaption{Key Kinematic Parameters of the Sample Stars \label{tab:orbit_short}}
\tablewidth{0pt}
\tabletypesize{\scriptsize}
\tablehead{
\colhead{ID} &
\colhead{DIST} &
\colhead{RV} &
\colhead{$U$} &
\colhead{$V$} &
\colhead{$W$} &
\colhead{$J_r$} &
\colhead{$J_\phi$} &
\colhead{$J_z$} &
\colhead{$E$} &
\colhead{$L_z$} &
\colhead{$e$} &
\colhead{$Z_{\rm max}$} \\
\colhead{} &
\colhead{(kpc)} &
\colhead{(km\,s$^{-1}$)} &
\colhead{(km\,s$^{-1}$)} &
\colhead{(km\,s$^{-1}$)} &
\colhead{(km\,s$^{-1}$)} &
\colhead{(kpc\,km\,s$^{-1}$)} &
\colhead{(kpc\,km\,s$^{-1}$)} &
\colhead{(kpc\,km\,s$^{-1}$)} &
\colhead{($10^{5}$ km$^{2}$\,s$^{-2}$)} &
\colhead{($10^{3}$ kpc\,km\,s$^{-1}$)} &
\colhead{} &
\colhead{(kpc)}
}
\startdata
J0858-0809 & $2.16^{+0.09}_{-0.09}$ & $168.11 \pm 0.13$ & $-152.55^{+2.73}_{-2.82}$ & $-68.59^{+2.52}_{-2.42}$ & $95.56^{+1.23}_{-1.18}$ & $282.37^{+11.52}_{-9.87}$ & $1896.92^{+47.45}_{-45.25}$ & $144.85^{+3.61}_{-3.46}$ & $-1.420^{+0.014}_{-0.013}$ & $1.897^{+0.047}_{-0.045}$ & $0.409^{+0.003}_{-0.002}$ & $4.77^{+0.18}_{-0.17}$ \\
J1930-1739 & $3.92^{+0.18}_{-0.17}$ & $104.55 \pm 0.08$ & $56.60^{+1.62}_{-1.71}$ & $39.29^{+0.28}_{-0.27}$ & $-143.50^{+4.99}_{-5.25}$ & $188.61^{+15.20}_{-12.71}$ & $1399.91^{+42.26}_{-44.32}$ & $172.43^{+9.83}_{-9.55}$ & $-1.562^{+0.003}_{-0.003}$ & $1.400^{+0.042}_{-0.044}$ & $0.393^{+0.016}_{-0.015}$ & $4.08^{+0.14}_{-0.21}$ \\
J1944-0837 & $3.04^{+0.12}_{-0.12}$ & $40.64 \pm 0.14$ & $73.53^{+1.57}_{-1.53}$ & $-95.55^{+4.35}_{-4.62}$ & $-102.24^{+3.43}_{-3.64}$ & $36.24^{+1.71}_{-1.74}$ & $948.33^{+33.32}_{-34.23}$ & $115.74^{+9.94}_{-8.93}$ & $-1.816^{+0.008}_{-0.008}$ & $0.948^{+0.033}_{-0.034}$ & $0.222^{+0.008}_{-0.008}$ & $2.19^{+0.10}_{-0.09}$ \\
J1950-1255 & $7.08^{+1.51}_{-1.19}$ & $-190.36 \pm 0.11$ & $-79.82^{+16.98}_{-13.35}$ & $-256.29^{+28.82}_{-36.69}$ & $32.80^{+4.83}_{-6.22}$ & $140.44^{+96.70}_{-34.32}$ & $-283.30^{+91.49}_{-34.63}$ & $172.86^{+18.29}_{-21.40}$ & $-2.027^{+0.036}_{-0.009}$ & $-0.283^{+0.091}_{-0.035}$ & $0.638^{+0.130}_{-0.046}$ & $2.80^{+0.39}_{-0.14}$ \\
J1952-1836 & $4.06^{+0.61}_{-0.54}$ & $-163.85 \pm 0.09$ & $-164.30^{+3.16}_{-3.57}$ & $-209.77^{+20.30}_{-22.87}$ & $-140.15^{+26.99}_{-30.08}$ & $550.72^{+51.61}_{-100.36}$ & $-84.25^{+158.53}_{-156.88}$ & $161.72^{+68.63}_{-52.97}$ & $-1.801^{+0.023}_{-0.006}$ & $-0.084^{+0.159}_{-0.157}$ & $0.870^{+0.042}_{-0.064}$ & $3.72^{+0.98}_{-0.71}$ \\
J2001-1215 & $3.80^{+0.28}_{-0.26}$ & $-12.99 \pm 0.08$ & $30.40^{+3.03}_{-2.81}$ & $-173.09^{+11.42}_{-12.43}$ & $-117.22^{+8.33}_{-9.06}$ & $172.05^{+6.86}_{-8.38}$ & $404.91^{+68.62}_{-68.56}$ & $196.51^{+40.43}_{-32.40}$ & $-1.923^{+0.008}_{-0.006}$ & $0.405^{+0.069}_{-0.069}$ & $0.572^{+0.026}_{-0.027}$ & $3.37^{+0.38}_{-0.34}$ \\
J2020-1002 & $6.29^{+0.67}_{-0.63}$ & $-216.74 \pm 0.10$ & $-59.73^{+10.97}_{-10.39}$ & $-201.43^{+9.08}_{-9.74}$ & $167.76^{+8.25}_{-7.83}$ & $28.95^{+32.36}_{-8.04}$ & $-64.04^{+47.31}_{-33.47}$ & $887.77^{+4.76}_{-3.92}$ & $-1.820^{+0.014}_{-0.005}$ & $-0.064^{+0.047}_{-0.033}$ & $0.295^{+0.060}_{-0.020}$ & $5.86^{+0.21}_{-0.07}$ \\
J2023-0815 & $2.21^{+0.07}_{-0.07}$ & $-176.08 \pm 0.14$ & $-139.98^{+0.34}_{-0.33}$ & $-356.45^{+7.78}_{-7.77}$ & $-287.83^{+10.67}_{-10.67}$ & $493.30^{+86.18}_{-66.37}$ & $-914.24^{+51.75}_{-50.81}$ & $1027.56^{+40.67}_{-41.83}$ & $-1.359^{+0.038}_{-0.036}$ & $-0.914^{+0.052}_{-0.051}$ & $0.550^{+0.023}_{-0.020}$ & $15.32^{+1.46}_{-1.31}$ \\
J2026-0047 & $1.46^{+0.04}_{-0.04}$ & $-232.27 \pm 0.10$ & $-124.08^{+0.85}_{-0.82}$ & $-216.43^{+1.68}_{-1.79}$ & $26.71^{+1.46}_{-1.54}$ & $689.68^{+4.08}_{-3.84}$ & $77.91^{+15.20}_{-16.04}$ & $22.54^{+0.66}_{-0.64}$ & $-1.800^{+0.003}_{-0.003}$ & $0.078^{+0.015}_{-0.016}$ & $0.957^{+0.010}_{-0.002}$ & $0.90^{+0.13}_{-0.03}$ \\
J2038-0023 & $6.00^{+0.48}_{-0.45}$ & $-323.10 \pm 0.11$ & $-35.88^{+13.74}_{-12.74}$ & $-551.15^{+25.59}_{-27.25}$ & $-147.17^{+20.97}_{-22.44}$ & $508.54^{+433.84}_{-250.55}$ & $-1509.11^{+26.20}_{-6.36}$ & $518.01^{+194.59}_{-145.60}$ & $-1.350^{+0.126}_{-0.102}$ & $-1.509^{+0.026}_{-0.006}$ & $0.515^{+0.105}_{-0.111}$ & $11.53^{+6.12}_{-4.47}$ \\
J2043-1305 & $1.35^{+0.03}_{-0.03}$ & $-128.18 \pm 0.11$ & $-173.34^{+1.94}_{-1.93}$ & $-171.45^{+2.62}_{-2.62}$ & $-150.78^{+5.09}_{-5.18}$ & $663.71^{+16.07}_{-15.38}$ & $411.71^{+24.57}_{-24.55}$ & $282.79^{+25.93}_{-24.13}$ & $-1.598^{+0.008}_{-0.008}$ & $0.412^{+0.025}_{-0.025}$ & $0.796^{+0.003}_{-0.004}$ & $8.31^{+0.39}_{-1.56}$ \\
J2140-1227 & $2.64^{+0.21}_{-0.20}$ & $-130.97 \pm 0.13$ & $113.47^{+14.86}_{-13.91}$ & $-359.52^{+22.22}_{-23.76}$ & $28.52^{+4.58}_{-4.93}$ & $365.74^{+22.05}_{-7.50}$ & $-644.90^{+113.82}_{-113.20}$ & $65.56^{+11.12}_{-6.09}$ & $-1.760^{+0.044}_{-0.034}$ & $-0.645^{+0.114}_{-0.113}$ & $0.647^{+0.032}_{-0.018}$ & $2.72^{+0.06}_{-0.32}$ \\
J2326-0159 & $4.14^{+0.28}_{-0.27}$ & $-217.82 \pm 0.13$ & $162.93^{+12.56}_{-11.76}$ & $-403.34^{+18.67}_{-19.90}$ & $23.34^{+10.40}_{-11.09}$ & $598.95^{+47.75}_{-25.08}$ & $-910.65^{+97.67}_{-98.79}$ & $197.29^{+79.97}_{-50.23}$ & $-1.513^{+0.059}_{-0.049}$ & $-0.911^{+0.098}_{-0.099}$ & $0.687^{+0.019}_{-0.018}$ & $5.65^{+2.03}_{-1.08}$ \\
\enddata
\tablecomments{
Only representative kinematic quantities directly discussed in the text are shown here. The full orbital parameter catalog, including all derived orbital and velocity quantities, is available in machine-readable form. Uncertainties correspond to the 16th and 84th percentiles.
}
\end{deluxetable*}
\end{longrotatetable}

\section{Kinematic Properties} \label{Kinematic}

In this study, astrometric parameters are adopted from {\it Gaia} DR3\footnote{\url{https://gea.esac.esa.int/archive/}} \citep{2023A&A...674A...1G}. To derive the kinematic properties of the sample stars, we combine these astrometric data with radial velocities measured from our high-resolution CFHT/ESPaDOnS spectra and the Bailer-Jones geometric distances \citep{2021AJ....161..147B}.

Orbital calculations are performed using the \texttt{galpy} Python package \citep{2015ApJS..216...29B}. The derived dynamical quantities include the Galactocentric cylindrical coordinates $(R,\phi,Z)$, Galactic velocity components ($U$, $V$, $W$), and cylindrical velocity components ($v_r$, $v_\phi$, $v_z$), orbital eccentricity ($e$), pericenter and apocenter distances ($r_{\rm peri}$ and $r_{\rm apo}$), maximum vertical distance from the Galactic plane ($Z_{\rm max}$), orbital energy ($E$), angular momentum about the Galactic center ($L_z$), and orbital actions ($J_r$, $J_\phi$, $J_z$). Positive $U$, $V$, and $W$ correspond to motions toward the Galactic center, in the direction of Galactic rotation, and toward the north Galactic pole, respectively. The cylindrical velocity components $(v_r, v_\phi, v_z)$ follow the standard Galactocentric cylindrical convention. Following the convention adopted in \citet{2024ApJ...966..174Z}, retrograde motions correspond to negative values of $L_z$ and $J_\phi$. We adopt Solar motion values of ($U_\odot$, $V_\odot$, $W_\odot$)\,$= (7.01,\,10.13,\,4.95)$\,km\,s$^{-1}$ \citep{2015MNRAS.449..162H}, and a circular velocity at the Solar radius of $V_c(R_0)=234.04$\,km\,s$^{-1}$ \citep{2023ApJ...946...73Z}. The solar position is assumed to be $R_0 = 8.178$\,kpc \citep{2019A&A...625L..10G} from the Galactic center and $Z_0 = 25$\,pc above the Galactic mid-plane \citep{2016ARA&A..54..529B}. Orbital parameters are computed by integrating the stellar orbits for a total of 1\,Gyr, with 5000 equally spaced time steps, within an axisymmetric Milky Way potential taken from \citet{2017MNRAS.465...76M}. The adopted astrometric parameters and the derived dynamical quantities are summarized in Table\,\ref{tab:orbit_short}.

Uncertainties in the dynamical parameters are estimated via Monte Carlo (MC) simulations with 5000 realizations for each star. Measurement uncertainties in proper motions and radial velocities are assumed to follow Gaussian distributions. For distances, whose uncertainties are non-Gaussian, we approximate the error distribution as log-normal using the reported 16th, 50th, and 84th percentiles from \citet{2021AJ....161..147B}. The uncertainties of the derived orbital parameters are then taken as the 16th and 84th percentiles of the resulting MC distributions. These uncertainties are reported alongside the corresponding dynamical parameters in Table\,\ref{tab:orbit_short}.



\section{Discussion} \label{Discussion}




\subsection{Overall Abundance Interpretation} 


We compare the chemical abundance trends of our sample stars with those of literature samples, shown as background points in Figures~\ref{fig:Li_C_ab}--\ref{fig:Sr_Ba}. Below, we first provide an overview of the element-by-element abundance trends for the full sample, and then highlight several stars with distinctive abundance patterns in more detail. The error bars shown for our program stars in Figures~\ref{fig:Li_C_ab}--\ref{fig:Sr_Ba} represent the total abundance uncertainties. For the literature comparison samples, only detections with reported abundance measurements are included, while upper limits are excluded. To avoid visual crowding, uncertainties for the literature samples are not shown in Figures~\ref{fig:Li_C_ab}, \ref{fig:abundance} and \ref{fig:Sr_Ba}. In Figure~\ref{fig:K_abundance_with_SAGA}, error bars are plotted when available and follow the uncertainties reported in the source literature or compiled in the SAGA database.

\subsubsection{General Abundance Trends}
\label{discussion:general_abundance_trends}

\begin{figure*}
\centering
\includegraphics[width=0.45\linewidth]{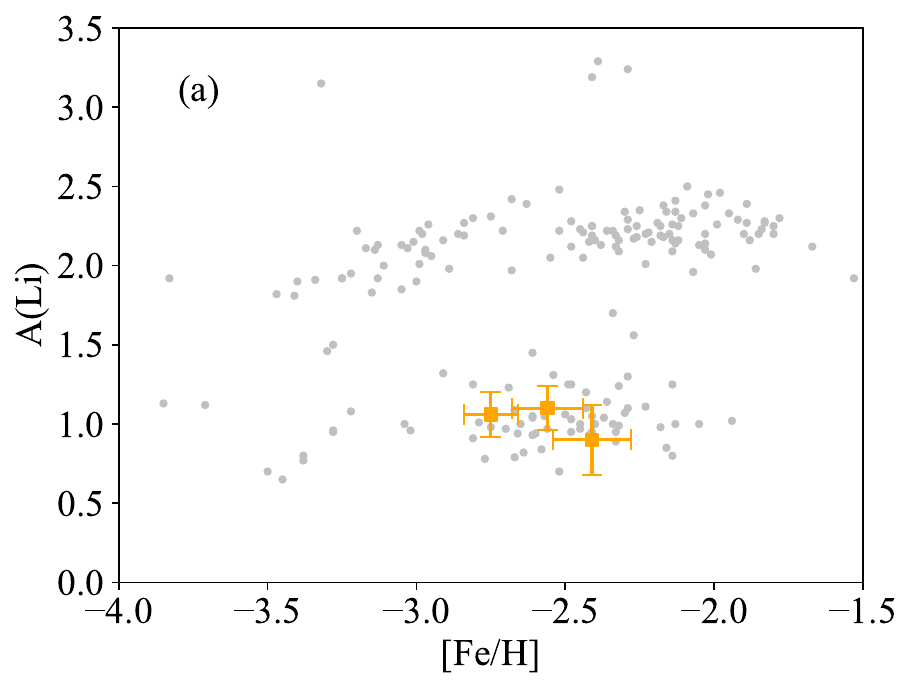}
\includegraphics[width=0.46\linewidth]{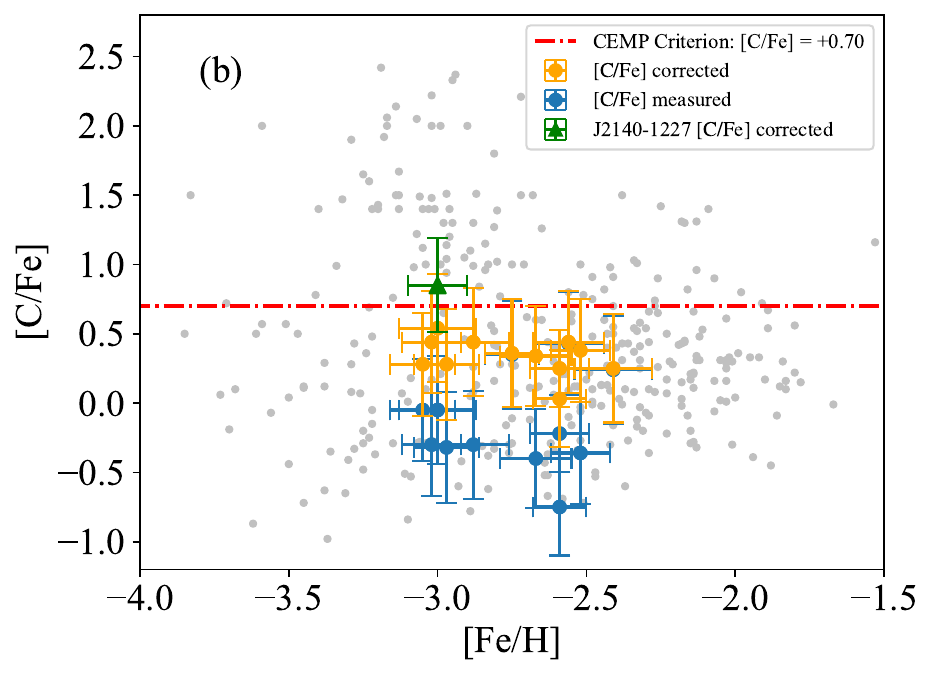}
\caption{(a) Lithium abundance, $A(\mathrm{Li})$, as a function of metallicity. Orange symbols denote stars in our sample with measured lithium abundances. (b) Carbon abundance as a function of metallicity. Blue and orange symbols represent the measured and evolution-corrected carbon abundances, respectively, with the latter calculated following \citet{2014ApJ...797...21P}. The green triangle highlights the corrected carbon abundance of J2140-1227. The dash-dotted line marks the adopted CEMP threshold, $\mathrm{[C/Fe]}_{\rm corr}=+0.70$, following \citet{2007ApJ...655..492A}. In both panels, gray background points are taken from \citet{2022ApJ...931..147L}, while the horizontal and vertical error bars for the program stars represent the $\mathrm{[Fe/H]}$ uncertainties reported in Table~\ref{tab: abundance} and the corresponding total abundance uncertainties, respectively.}
\label{fig:Li_C_ab}
\end{figure*}

Overall, the abundance trends of our sample stars are broadly consistent with those of the comparison sample of VMP and EMP stars compiled by \citet{2022ApJ...931..147L}. The detected lithium abundances are confined to the less evolved stars with $\log g>2.0$, consistent with the lithium depletion expected after the first dredge-up. After applying the evolutionary corrections provided by \citet{2014ApJ...797...21P}, most stars remain carbon-normal. For classification, we adopt $\mathrm{[C/Fe]}_{\rm corr}>+0.70$, based on the revised CEMP definition of \citet{2007ApJ...655..492A}, rather than the earlier threshold of $\mathrm{[C/Fe]}>+1.0$ adopted by \citet{2005ARA&A..43..531B}. Only J2140-1227 satisfies the adopted criterion, as highlighted in the right panel of Figure~\ref{fig:Li_C_ab}. For the light and $\alpha$ elements, most stars show the typical enhancement relative to Fe observed in metal-poor halo stars, reflecting the dominant contribution of core-collapse supernovae to early chemical enrichment. The iron-peak elements also generally follow the trends defined by the comparison sample, whereas the neutron-capture elements exhibit substantially larger star-to-star scatter, consistent with the rarity and inhomogeneous nature of early neutron-capture enrichment.

Against these generally consistent abundance trends, several stars stand out as chemically peculiar. The most chemically complex object is J2140-1227, which exhibits coherent abundance anomalies across multiple nucleosynthetic groups, including C, Na, several iron-peak elements, and light neutron-capture elements. J2001-1215 shows a different type of peculiarity, namely a strong K enhancement, while J2023-0815 is distinguished by its exceptionally low neutron-capture abundances. In addition, three stars in our sample, J2001-1215, J2038-0023, and J2326-0159, satisfy the $r$-II classification criterion of \citet{2020ApJS..249...30H}, with $\mathrm{[Eu/Fe]}>0.70$ and $\mathrm{[Ba/Eu]}<0.00$. In the following subsection, comparisons with previous high-resolution analyses of J2140-1227, J2038-0023, and J2326-0159 are used to validate our atmospheric parameters and abundance measurements. J2001-1215 and J2023-0815 are subsequently discussed in dedicated subsections, with emphasis on their potassium enhancement and possible accreted origin, respectively.

\begin{figure*}
\centering
\includegraphics[width=0.95\linewidth]{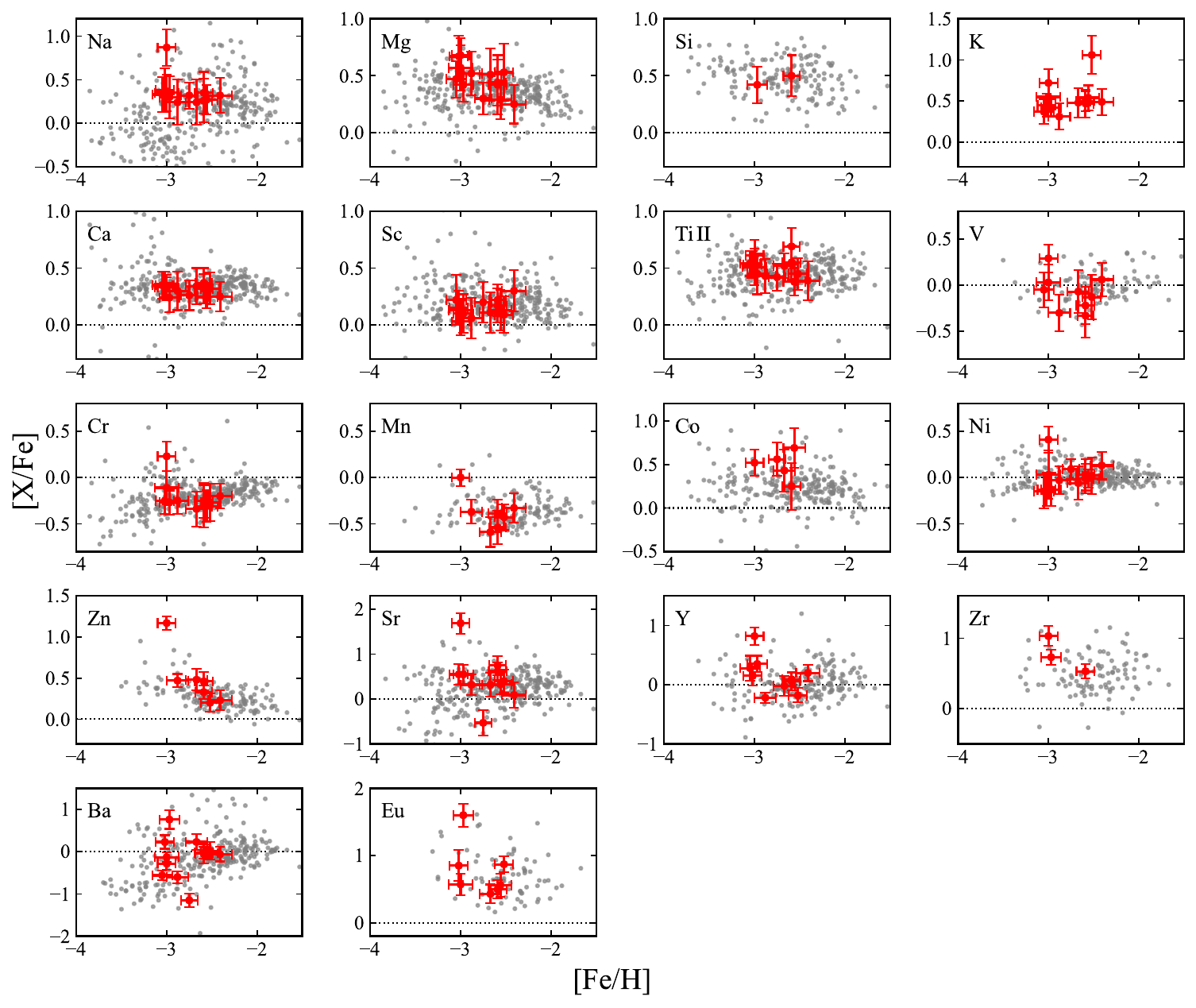}
\caption{Abundance ratios as a function of metallicity for 18 elements. Red symbols represent stars in our sample, while gray background symbols denote comparison stars from \citet{2022ApJ...931..147L}. For the program stars, the horizontal error bars represent the $\mathrm{[Fe/H]}$ uncertainties, while the vertical error bars represent the total abundance uncertainties in $\mathrm{[X/Fe]}$. The comparison sample does not provide potassium abundances; therefore, no background data are shown in the K panel. For titanium, only Ti\,{\sc ii} abundances are available in the comparison sample, and the Ti panel accordingly displays Ti\,{\sc ii} abundances for both our sample and the comparison stars.}
\label{fig:abundance}
\end{figure*}

\subsubsection{Validation against Previous High-Resolution Studies}
\label{discussion:literature_validation}


J2140-1227 provides one of the most useful external consistency checks for the present analysis. Among the stars in common with previous high-resolution work, it is the object for which both the atmospheric-parameter determination method and the resulting parameters are most similar to those adopted in the literature. The abundance pattern derived in this work is also in excellent agreement with the high-resolution analysis of \citet{2024Dornbusch}. In that study, J2140-1227 was classified as a limited-$r$ star \citep{2018ARNPS..68..237F}, characterized by strong enhancement of light neutron-capture elements such as Sr, Y, and Zr, but without a corresponding strong enhancement of heavier neutron-capture elements such as Ba and Eu. We recover the same qualitative pattern from the CFHT/ESPaDOnS spectrum: J2140-1227 is strongly enhanced in Sr, Y, and Zr ([Sr/Fe] = 1.69, [Y/Fe] = 0.82, and [Zr/Fe] = 1.03), while its Ba abundance remains much lower than the light neutron-capture abundances ($\rm [Ba/Fe] = -0.28$) and Eu cannot be robustly constrained from our spectrum. Beyond the light neutron-capture enhancement, the C, Na, and iron-peak peculiarities reported by \citet{2024Dornbusch} are also recovered in our analysis. Taken together, this agreement with previous high-resolution results supports the reliability of our homogeneous atmospheric-parameter scale and abundance measurements.

J2038-0023 provides an additional check on our heavy-element abundance measurements. This star was previously identified by \citet{2017ApJ...844...18P} as a strongly $r$-process-enhanced star, with $\mathrm{[Eu/Fe]}=1.64$ and $\mathrm{[Ba/Eu]}=-0.81$. The atmospheric parameters adopted in the present work are in good agreement with those of that study, and our derived neutron-capture abundances are likewise highly consistent. We obtain $\mathrm{[Eu/Fe]}=1.60$ and $\mathrm{[Ba/Eu]}=-0.87$, confirming J2038-0023 as an $r$-II star within our homogeneous analysis. Since both Ba and Eu are clearly detected in this star, the agreement with previous high-resolution spectroscopy provides an important consistency check on our heavy-element abundance determination for strongly $r$-process-enhanced EMP stars.

J2326-0159 requires more careful discussion because its europium enhancement lies close to the adopted boundary between the $r$-I ($0.30\leq\mathrm{[Eu/Fe]}\leq0.70$) and $r$-II ($\mathrm{[Eu/Fe]}>0.70$) classes, both of which require $\mathrm{[Ba/Eu]}<0.00$ \citep{2020ApJS..249...30H}. This star was first analyzed in the high-resolution study of \citet{Thanathibodee2016}, who derived $\mathrm{[Eu/Fe]}=0.73$, which satisfies the $r$-II criterion adopted in the present work. It was later classified as an $r$-I star by \citet{2018ApJ...858...92H}, who reported a slightly lower europium enhancement of $\mathrm{[Eu/Fe]}=0.69$. Both studies used a primarily spectroscopic parameter determination, together with the temperature correction procedure of \citet{2013ApJ...769...57F}. \citet{Thanathibodee2016} derived $T_{\rm eff}=4477$\,K, $\log g=0.70$, $\mathrm{[Fe/H]}=-3.05$, and $v_{\rm mic}=2.70$\,km\,s$^{-1}$, while \citet{2018ApJ...858...92H} adopted $T_{\rm eff}=4360$\,K, $\log g=0.45$, $\mathrm{[Fe/H]}=-3.13$, and $v_{\rm mic}=2.40$\,km\,s$^{-1}$. In the present work, we adopt $T_{\rm eff}=4477$\,K, $\log g=0.90$, $\mathrm{[Fe/H]}=-3.02$, and $v_{\rm mic}=1.91$\,km\,s$^{-1}$, based on photometric temperatures and {\it Gaia}~DR3 distance-based surface gravities.

Under our adopted parameters, we derive $\mathrm{[Eu/Fe]}=0.87$ and $\mathrm{[Ba/Eu]}=-0.62$, placing J2326-0159 in the $r$-II regime. The higher europium abundance mainly reflects the combined effect of the different atmospheric parameters adopted in this work. Our sensitivity analysis shows that increasing $T_{\rm eff}$ by 100\,K raises the derived Eu abundance by 0.05\,dex, increasing $\log g$ by 0.3\,dex raises it by 0.03\,dex, and increasing $v_{\rm mic}$ by 0.2\,km\,s$^{-1}$ lowers it by 0.02\,dex. Compared with \citet{Thanathibodee2016} and \citet{2018ApJ...858...92H}, our adopted effective temperature is identical to the former and higher than the latter by 117\,K, while our adopted surface gravity is higher by 0.20 and 0.45\,dex, respectively, and our adopted microturbulent velocity is lower by 0.79 and 0.49\,km\,s$^{-1}$, respectively. These parameter differences act in the direction of increasing the derived Eu abundance in our analysis. Although the abundance response to each parameter is modest, their combined effect is sufficient to shift J2326-0159 across the adopted $r$-I/$r$-II boundary because the literature values already place the star very close to this threshold. Thus, the available high-resolution analyses indicate a consistent heavy-element abundance pattern for J2326-0159, while the different classifications mainly arise from modest systematic differences in the adopted atmospheric parameters and the discrete nature of the $r$-I/$r$-II classification criterion.

Taken together, the comparisons with previous high-resolution analyses support the reliability of our atmospheric parameter scale and abundance measurements. The results for J2140-1227 and J2038-0023 show close agreement with the literature, while J2326-0159 illustrates how modest systematic differences in atmospheric parameters can shift a star across the discrete $r$-I/$r$-II classification boundary without implying a substantial disagreement in its underlying heavy-element abundance pattern.

\subsubsection{J2001-1215: a potassium-enhanced $r$-II star} \label{discussion: J2001-1215}

J2001-1215 is newly identified in this work as a chemically peculiar star with both a strong potassium enhancement and an $r$-process enhancement. It has $\mathrm{[Eu/Fe]}=0.87$ and $\mathrm{[Ba/Eu]}=-0.85$, placing it in the $r$-II regime and indicating a dominant $r$-process contribution to its heavy-element abundance pattern. At the same time, it exhibits a remarkably high potassium abundance, $\mathrm{[K/Fe]}=1.06$ under LTE, while no comparable anomaly is detected among the other measurable light elements. A Na abundance, which would provide an important comparison as another odd-$Z$ element, could not be derived because the Na\,{\sc i}\,D doublet is severely contaminated by interstellar Na absorption. We also note that the K\,{\sc i}\,$\lambda$7664 and K\,{\sc i}\,$\lambda$7698 resonance lines used for the K abundance determination may potentially be affected by telluric water-vapor absorption. To assess the reliability of the potassium enhancement, we carefully inspected the spectral region between 7600 and 7740\,\AA. The K\,{\sc i}\,$\lambda$7664 line is blended with strong telluric absorption and cannot be reliably identified, whereas the K\,{\sc i}\,$\lambda$7698 line lies cleanly between two groups of telluric features and is not affected. The potassium abundance is therefore derived solely from the uncontaminated K\,{\sc i}\,$\lambda$7698 line, for which we measure an equivalent width of 108.3\,m\AA\ in the CFHT/ESPaDOnS spectrum. The observed Na\,{\sc i}\,D and K\,{\sc i} resonance-line regions are shown in Appendix Figure~\ref{fig:NaK_region}. Furthermore, a direct comparison of the K\,{\sc i}\,$\lambda$7698 region in J2001-1215 with stars of similar atmospheric parameters confirms the intrinsic strength of its potassium absorption.

\begin{figure}
\centering
\includegraphics[width=1.0\linewidth]{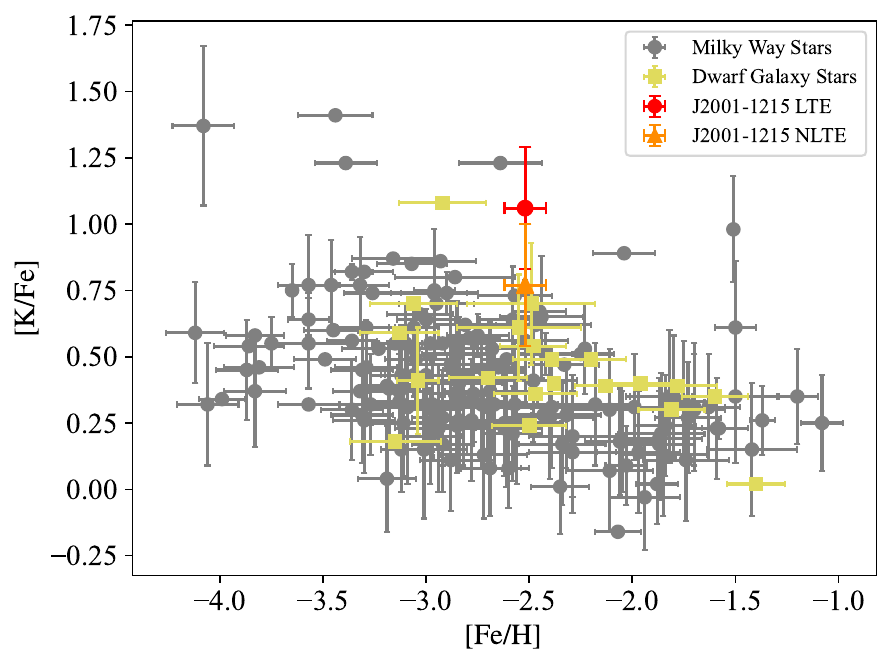}
\caption{The [K/Fe] ratio as a function of [Fe/H]. The comparison sample is compiled from the SAGA database \citep{2008PASJ...60.1159S} and likely includes a mixture of LTE and NLTE abundance determinations from heterogeneous literature sources. Gray and yellow symbols denote stars belonging to the Milky Way and dwarf galaxies, respectively. The red dot and orange triangle indicate the LTE and NLTE potassium abundances of J2001-1215 derived from the CFHT/ESPaDOnS spectrum.}
\label{fig:K_abundance_with_SAGA}
\end{figure}

Since J2001-1215 exhibits an unusually high potassium abundance, we compare its [K/Fe] ratio with stars in the SAGA database\footnote{\url{http://sagadatabase.jp/}} \citep[Figure~\ref{fig:K_abundance_with_SAGA}]{2008PASJ...60.1159S} to assess the rarity of such enhancement among metal-poor stellar populations. The SAGA database compiles measurements from heterogeneous literature sources, and the listed potassium abundances likely include a mixture of LTE and NLTE determinations. We therefore use this comparison primarily as a qualitative reference. Moreover, because NLTE corrections for the K\,{\sc i} resonance lines are usually large and negative, an LTE-only comparison could overestimate the significance of the potassium enhancement in J2001-1215. We therefore compute an NLTE correction for its K abundance to enable a more robust comparison.

The NLTE calculation is performed using the atomic model of \citet{2006Zhang} and the Spectrum Investigation Utility (SIU; \citealt{1991PhDThesis}). Using the same measured EW of the K\,{\sc i}\,$\lambda$7698 line from the CFHT/ESPaDOnS spectrum, we derive the potassium abundance under both LTE and NLTE assumptions. The difference between these two abundances gives an NLTE correction of $-0.29$\,dex, yielding $\mathrm{[K/Fe]_{NLTE}=0.77}$. J2001-1215 therefore remains K-enhanced after the NLTE correction, indicating that the enhancement is not an LTE artifact. The abundance uncertainty does not include possible systematic uncertainties associated with the NLTE calculation itself. Consistent with the treatment of other single-line abundances in this work, we adopt the Fe\,{\sc i} line-to-line scatter as the statistical uncertainty.


As a supporting consistency check on the CFHT/ESPaDOnS results, we also examined a follow-up Magellan/MIKE spectrum. The Magellan/MIKE observation was obtained using the 0.7'' slit, providing wavelength coverage from $\sim3300$ to $9400$\,\AA\ with a resolving power of $R\sim35,000$ in the blue arm. The spectrum reaches a signal-to-noise ratio exceeding 500 around the K\,{\sc i}\,$\lambda7698$ region. Using the same stellar atmospheric parameters and abundance-analysis framework adopted for the CFHT/ESPaDOnS analysis, the Magellan/MIKE spectrum yields $\rm EW=113.8$\,m\AA\ and a consistent potassium abundance of $\rm [K/Fe]=1.15$, independently confirming the intrinsic strength of the K\,{\sc i}\,$\lambda7698$ feature. The Magellan/MIKE spectrum also yields $\mathrm{[Eu/Fe]}=0.75$, supporting the $r$-process-enhanced nature inferred from the CFHT/ESPaDOnS spectrum. A comprehensive analysis of the Magellan/MIKE spectrum, including measurements of additional neutron-capture elements required to characterize the detailed r-process abundance pattern, is beyond the scope of this work and will be presented in a future study.

Although the potassium enhancement is therefore robust observationally, its nucleosynthetic origin remains unclear. Current Galactic chemical evolution models do not yet provide reliable predictions for potassium abundances \citep{2020ApJ...900..179K}, complicating the interpretation. Theoretically, potassium production may be enhanced by stellar rotation \citep{2018MNRAS.476.3432P} or by O--C shell mergers during hydrostatic burning \citep{2018MNRAS.474L...1R}. These scenarios, however, are also expected to produce scandium enhancements, which are not observed in J2001-1215, highlighting the limitations of current nucleosynthetic models in explaining isolated potassium enhancement.

More recently, \citet{2025ApJ...992..215I} analyzed seven extremely metal-poor stars and reported a small intrinsic scatter in [K/Fe] and [K/Ca], in stark contrast to the $\sim$0.7\,dex scatter observed in [Na/Mg]. They argued that potassium production in massive stars or their supernovae may be largely decoupled from the nucleosynthetic pathways governing Na and Mg. This scenario is qualitatively consistent with the isolated potassium enhancement observed in J2001-1215. Nevertheless, the limited size of current EMP samples precludes firm conclusions, and a larger, homogeneous dataset will be essential for elucidating the origin of potassium-enhanced stars at the lowest metallicities.

\subsubsection{J2023-0815: accreted dwarf galaxy remnant}\label{J2023-0815}

J2023-0815 exhibits the lowest neutron-capture element abundances among our sample stars. When compared with the larger compilation of very metal-poor (VMP) and extremely metal-poor (EMP) stars presented by \citet{2022ApJ...931..147L}, its neutron-capture abundances are clearly lower than those of the majority of stars at similar metallicities. Such extremely low neutron-capture abundances are typically associated with environments in which star formation was inefficient and quenched at early times, most notably low-mass systems such as ultra-faint dwarf (UFD) galaxies \citep{2010Natur.464...72F,2018ARNPS..68..237F}.

\begin{figure}
\centering
\includegraphics[width=1.0\linewidth]{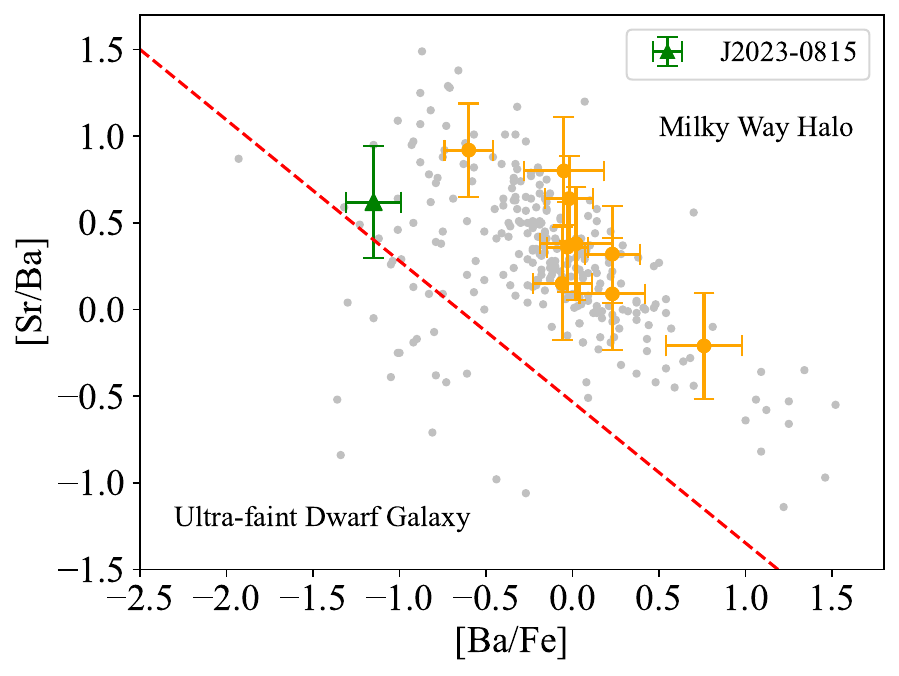}
\caption{The [Sr/Ba] versus [Ba/Fe] abundance ratios for the sample stars analyzed in this work (orange symbols). The background gray dots represent comparison stars from \citet{2022ApJ...931..147L}. The green triangle highlights J2023-0815, which is discussed in Sec.\,\ref{J2023-0815} as a chemically peculiar object. The red dashed line indicates the empirical separation between Milky Way halo stars and stars associated with ultra-faint dwarf galaxies adopted in previous studies \citep{2024MNRAS.530.4712A,2025AJ....169..279O}.}
\label{fig:Sr_Ba}
\end{figure}

In recent studies, the position of stars in the [Sr/Ba] versus [Ba/Fe] plane has been widely used to investigate their formation environments, particularly to identify stars of possible UFD origin \citep{2024MNRAS.530.4712A,2025AJ....169..279O}. In Figure~\ref{fig:Sr_Ba}, we present the locations of our sample stars in this diagnostic plane. Amongst them, J2023-0815 lies closest to the empirical boundary separating Milky Way halo stars from those associated with UFD systems, suggesting a high probability that this star originated in a low-mass galaxy. This interpretation is further supported by its orbital properties (Table~\ref{tab:orbit_short}), including its relatively high orbital energy, large $Z_{\max}$, and retrograde motion, all of which are characteristic of accreted halo stars.

Although J2023-0815 has previously been analyzed in the literature \citep{2020ApJ...898..150E}, the substantial differences in adopted atmospheric parameters lead to discrepancies in the derived abundances, particularly for Sr, as the Sr\,{\sc ii}\,$\lambda$4077\,\AA\ line is highly sensitive to stellar parameters. Nevertheless, even when adopting the abundances reported by \citet{2020ApJ...898..150E} ($\rm [Sr/Fe]=-1.17$ and $\rm [Ba/Fe]=-0.95$), the resulting ratios ($\rm [Sr/Ba]=-0.22$ and $\rm [Ba/Fe]=-0.95$) place J2023-0815 firmly within the region associated with UFD-origin stars in the [Sr/Ba] versus [Ba/Fe] plane. Therefore, both our analysis and previous results consistently support an accretion origin for J2023-0815.

\subsection{Inspection with dynamically tagged groups}

Although numerous studies have attempted to identify dynamically tagged groups (DTGs) in the Milky Way, the criteria used to define such structures are not fully consistent across the literature. In this work, we adopt the classification scheme presented by \citet{2024ApJ...966..174Z} to investigate the possible associations between our sample stars and several major Galactic substructures in the Milky Way, including Gaia--Sausage--Enceladus (GES), Sequoia, Thamnos, the Helmi stream, Wukong, Pontus, and the very metal-poor disk. We follow the dynamical selection regions defined by \citet{2024ApJ...966..174Z} when assessing these possible associations.

\begin{figure*}
    \centering
    \includegraphics[width=0.9\linewidth]{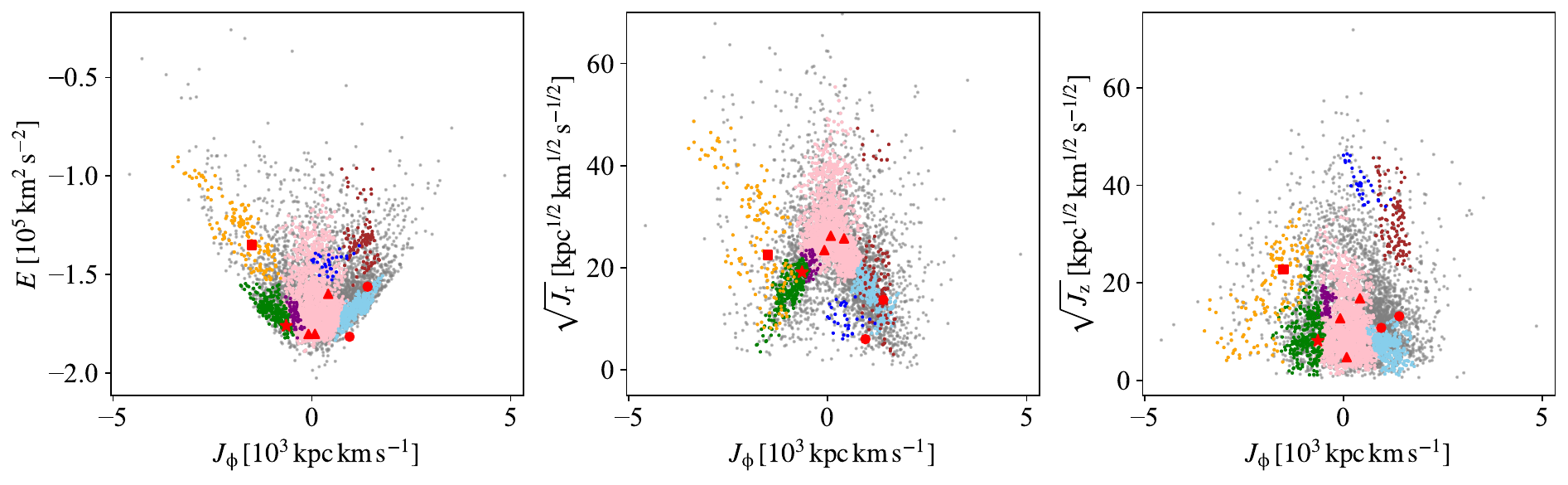}
    \caption{The distributions of our sample stars associated with known substructures in energy and action phase space. Red symbols highlight stars meeting specific substructure criteria: Sequoia (red square), Thamnos (red star), and Gaia--Sausage--Enceladus (GES; red triangles). Two stars (red dots) fall in the thick-disk region of the Toomre diagram but do not fully satisfy the selection criteria of \citet{2024ApJ...966..174Z}. The background stars are taken from \citet{2024ApJ...966..174Z}, with different colors representing difference substructures. The gray dots represent Milky Way halo stars, while orange, green, purple, pink, blue, brown and light blue represent Sequoia, Thamnos, Pontus, GES, Wukong, Helmi stream and VMP thin disk, respectively.}
    \label{fig:DTG}
\end{figure*}

High-resolution spectroscopic follow-up observations for stars associated with these Galactic substructures are currently limited, typically comprising only a few tens of stars (see Table~D1 in \citet{2024ApJ...966..174Z}), and in most cases do not extend to the very metal-poor regime (generally $\mathrm{[Fe/H] > -2.5}$). Our sample therefore extends the chemical coverage of these Galactic substructures down to $\mathrm{[Fe/H] \sim -3.0}$, providing a valuable opportunity to trace their early chemical evolution and to probe the nature of their progenitor systems at the lowest metallicities.

By comparing the orbital energy and action distributions of our sample stars with the dynamical selection regions of individual Galactic substructures, five stars are identified as dynamically consistent with these systems. Specifically, J1952-1836, J2026-0047, and J2043-1305 are associated with GES; J2140-1227 with Thamnos; and J2038-0023 with Sequoia. Their locations in the $E$--$J_{\phi}$, $\sqrt{J_{\rm r}}$--$J_{\phi}$, and $\sqrt{J_{\rm z}}$--$J_{\phi}$ planes are shown in Figure~\ref{fig:DTG}, together with background stars from \citet{2024ApJ...966..174Z}.

The three stars dynamically associated with GES exhibit abundance patterns broadly consistent with those of typical metal-poor halo stars, with no distinctive chemical anomalies identified in the elements measured here. This is consistent with previous studies showing that GES stars largely overlap with the Milky Way halo in chemical-abundance space, especially at low metallicity \citep[e.g.,][]{2021ApJ...908L...8A,2022MNRAS.513.1557C,2024ApJ...966..174Z}. Among the three GES-associated stars in our sample, one has a measured europium abundance of $\mathrm{[Eu/Fe]}=0.56$, while Eu is not detected in the other two. Given the limited number of Eu detections and the detection threshold of our spectra, we do not draw a firm conclusion on the $r$-process enrichment level of the GES-associated stars in this sample.

J2140-1227, associated with Thamnos, is an extremely metal-poor star with a highly peculiar chemical abundance pattern, as discussed in Section~\ref{discussion:literature_validation}. Its metallicity places it among the most metal-poor Thamnos-associated stars reported to date \citep{2010A&A...511L..10N,2011A&A...530A..15N,2020MNRAS.497.1236M,2024ApJ...966..174Z}. Existing Thamnos stars in the literature generally follow chemical trends similar to those of the Milky Way halo, and none exhibits a pattern comparable to that of J2140-1227. 

An r-process enhancement signature has also been reported for Sequoia stars \citep{2021ApJ...908L...8A}. Moreover, \citet{2024ApJ...966..174Z} found that stars associated with Sequoia and having orbital energies $E > -1.35 \times 10^{5}\,\mathrm{km^2\,s^{-2}}$ systematically exhibit $\mathrm{[Eu/Fe]}\sim0.7$. Notably, the Sequoia candidate identified in this work, J2038-0023, is strongly $r$-process enhanced ($\mathrm{[Eu/Fe]}=1.60$) and has an orbital energy of $E=-1.350^{+0.126}_{-0.102}\times10^{5}\,\mathrm{km^2\,s^{-2}}$, placing it at the boundary of the high-energy Sequoia region defined by \citet{2024ApJ...966..174Z}, with its uncertainty extending into this regime. Its strong $r$-process enhancement is therefore consistent with the trend identified for the high-energy component of Sequoia and suggests that this component may have experienced one or more prolific $r$-process enrichment events. Further investigation of stars in this region of action--energy space will be crucial for constraining the origin, frequency, and astrophysical site(s) of r-process nucleosynthesis in accreted Galactic substructures.

\begin{figure}
\centering
\includegraphics[width=1.0\linewidth]{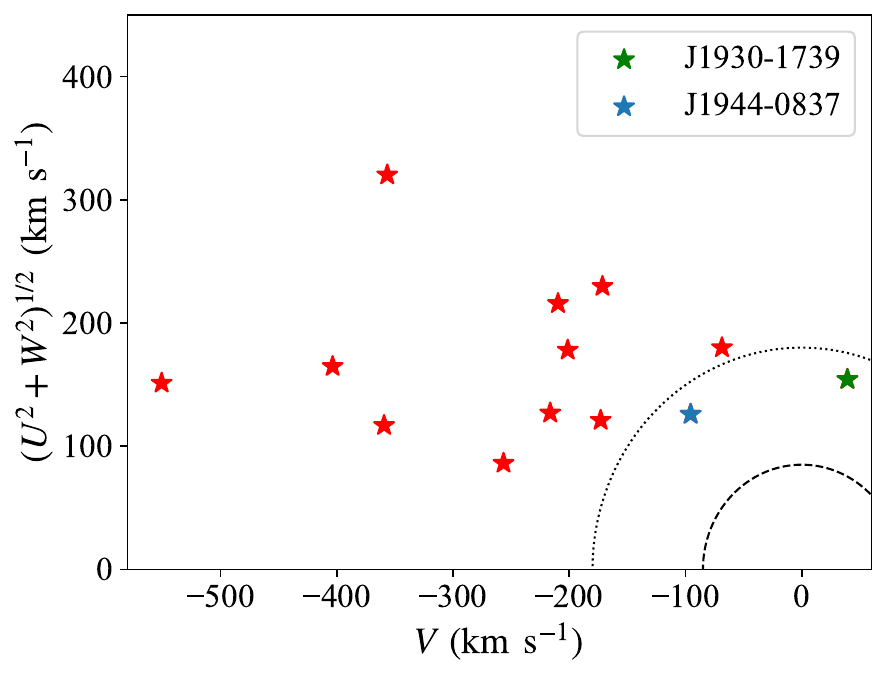}
\caption{Toomre diagram for the sample stars. J1930-1739 and J1944-0837 are highlighted with green and blue star symbols, respectively, while the other sample stars are shown as red star symbols. The inner dashed and outer dotted circles represent constant total space velocities relative to the local standard of rest, with $V_{\rm tot}=85$ and $180$\,km\,s$^{-1}$, respectively.}
\label{fig:Toomre diagram}
\end{figure}

In addition to the five stars dynamically consistent with three Galactic substructures, two stars, J1930-1739 and J1944-0837, fall in the thick-disk region of the Toomre diagram shown in Figure~\ref{fig:Toomre diagram}. However, we do not classify them as robust very metal-poor thick-disk members under the criteria of \citet{2024ApJ...966..174Z}. For J1930-1739, the relatively large maximum vertical distance ($Z_{\rm max} > 3$\,kpc) is not typical of the thick-disk population. In contrast, J1944-0837 has a low-eccentricity ($e < 0.25$), disk-like orbit, making it a plausible disk member; however, its eccentricity is below the selection range adopted by \citet{2024ApJ...966..174Z} for very metal-poor thick-disk stars. We therefore regard J1944-0837 as a possible disk member and J1930-1739 as a more tentative disk-like candidate.

Chemically, these two stars do not display pronounced differences from typical Milky Way halo stars in terms of their overall abundance patterns. Nevertheless, recent studies \citep[e.g.,][]{2022ApJ...936...78M} have reported metal-poor stars with accretion signatures occupying a similar region of kinematic space, suggesting that the metal-poor tail of the disk may comprise a mixture of in-situ and accreted populations. Expanding the sample of stars with both reliable dynamical information and detailed chemical abundances is therefore essential for statistically characterizing this metal-poor disk component and for constraining the early formation and assembly history of the Milky Way disk.

\raggedbottom
\section{Summary} \label{sec:summary}

In this work, we present high-resolution spectroscopic follow-up observations of 13 EMP candidates selected from our photometric metallicity catalog based on the SkyMapper Southern Survey. This study represents the first high-resolution validation of EMP candidates identified by our photometric pipeline, and it provides an important benchmark for the future application of our photometric metallicity. The main results and conclusions of this work are summarized as follows:

\begin{enumerate}
\item The SkyMapper-based narrow-band photometric selection is effective in identifying very metal-poor stars and retaining a substantial EMP fraction: four out of the thirteen candidates are confirmed as EMP stars, all targets remain very metal-poor with $\mathrm{[Fe/H]} < -2.4$. The spectroscopic metallicities are systematically higher than the SkyMapper photometric estimates by a mean offset of $0.44$\,dex, with a scatter of $0.19$\,dex, quantifying a modest metallicity-scale offset in the current photometric catalog that can be improved through future calibration.

\item Detailed chemical abundances for more than 20 elements are derived. We identify one newly recognized $r$-II star with significant potassium enhancement, J2001-1215. Through a homogeneous reanalysis of several stars previously studied in the literature, we further find that improved distance constraints from {\it Gaia} can lead to systematically different atmospheric parameters for VMP and EMP stars, particularly in surface gravity, which in turn affects the detailed element abundance measurements.

\item By jointly analyzing chemical abundances and dynamical properties, we identify several stars dynamically consistent with the Gaia--Sausage--Enceladus (GES), Thamnos, and Sequoia Galactic substructures. In particular, we find a chemically peculiar EMP star associated with Thamnos, as well as an r-process--enhanced star associated with Sequoia, providing further evidence for r-process enrichment in the high-energy component of the Sequoia system.
\end{enumerate}

Overall, this study demonstrates that wide-field narrow-band photometric surveys, combined with a photometric metallicity-selection framework calibrated against spectroscopic metallicities, provide an efficient approach for prioritizing VMP and EMP candidates for high-resolution follow-up. The resulting high-resolution spectra further enable detailed chemical-abundance analyses and chemo-dynamical studies, as illustrated by the identification of the potassium-enhanced r-II star J2001-1215 and by the association of several metal-poor stars with known Galactic substructures. The SkyMapper-based candidate catalog therefore provides a useful resource for targeted spectroscopic studies of metal-poor stars and their connection to the early assembly history of the Milky Way.

\section*{acknowledgements}

We sincerely thank the referee for constructive comments that helped improve the clarity and quality of this paper. This work is funded by the National Key R\&D Program of China (No. 2024YFA1611903).
Y.H. acknowledges the support from the National Science Foundation of China (NSFC grant No. 12422303), the Fundamental Research Funds for the Central Universities (grant Nos. 118900M122, E5EQ3301X2, and E4EQ3301X2), and the National Key R\&D Program of China (grant No. 2023YFA1608303). ZG thanks the support from FONDECYT Iniciacion project No. 11260176.
This work is also supported by the Strategic Priority Research Program of the Chinese Academy of Sciences (XDB0550103), the China--Chile Joint Research Fund (CCJRF No.~2301), and the Chinese Academy of Sciences South America Center for Astronomy (CASSACA) Key Research Project (E52H540301).

This paper is primarily based on high-resolution spectra obtained with the Canada--France--Hawaii Telescope (CFHT). Thanks to the TAP program to provide the opportunity for applying observation on CFHT. This work also includes data obtained with the 6.5\,m Magellan Telescopes at Las Campanas Observatory, Chile, using the MIKE spectrograph on the Magellan Clay telescope, under observing program CN2025B--62 allocated by the Chilean Telescope Allocation Committee (CNTAC). This work also makes use of data from the European Space Agency ({\it ESA}) mission {\it Gaia} (\url{https://www.cosmos.esa.int/Gaia}), processed by the {\it Gaia} Data Processing and Analysis Consortium (DPAC; \url{http://www.cosmos.esa.int/web/Gaia/dpac/consortium}).


\clearpage
\bibliographystyle{aasjournal}
\bibliography{HRS_CFHT}

\clearpage
\onecolumngrid
\appendix

\section{Color Transformation} \label{color transformation}

Because the APASS $V$-band photometry of several sample stars has relatively large uncertainties, we adopt SMSS $g$-band magnitudes and transform the dereddened $(g-K_{\rm s})$ colors into $(V-K_{\rm s})$. Effective temperatures are then derived from the resulting $(V-K_{\rm s})$ colors using the calibration in Table~3 of \citet{2005ApJ...626..465R}.

The color transformation is calibrated using standard field stars from \citet{2013AJ....146...88C}, cross-matched with the 2MASS catalog \citep{2006AJ....131.1163S} and SMSS DR2 \citep{2019PASA...36...33O} to obtain homogeneous $V$, $K_{\rm s}$, and $g$-band photometry. Reddening corrections are applied using $E(B-V)$ values from the two-dimensional dust map of \citet{1998ApJ...500..525S}, with the recalibration of \citet{2011ApJ...737..103S}. The adopted extinction coefficients are $R_V = 3.100$, $R_g = 3.407$ \citep{2019ApJS..243....7H}, and $R_{K_{\rm s}} = 0.329$ \citep{2004AJ....128.2144M}.

We fit the relation between the dereddened colors $(g-K_{\rm s})$ and $(V-K_{\rm s})$ using a third-order polynomial of the form

\begin{equation}
(V-K_{\rm s}) = a_0 + a_1 X + a_2 X^2 + a_3 X^3,
\end{equation}

where $X = (g-K_{\rm s})$. The fitting is performed iteratively with two rounds of $3\sigma$ clipping to remove outliers. The resulting calibration coefficients are listed in Table~\ref{Coefficients of Calibration}, while the fitting performance is shown in Figure~\ref{fig:color_transformation}. The final transformation yields a dispersion of $\sigma = 0.0146$\,mag. Combined with the typical photometric uncertainties in $(g-K_{\rm s})$ of 0.021--0.030\,mag and the intrinsic uncertainty of the adopted color--temperature relation (28\,K; \citealt{2005ApJ...626..465R}), the resulting uncertainties in the derived effective temperatures are estimated to be approximately 35--50\,K for all sample stars. The transformed colors and the corresponding first-iteration input effective temperatures, derived using the photometric metallicities listed in Table~\ref{tab:observation_and_photometry_details}, are summarized in Table~\ref{table:photometric_temperature}. The final adopted atmospheric parameters after the iterative analysis are listed in Table~\ref{Table of Stellar Atmospheric Parameters}.

\FloatBarrier
\setcounter{figure}{0}
\renewcommand{\thefigure}{A.\arabic{figure}}
\begin{figure}[!htbp]
\centering
\includegraphics[width=0.5\linewidth]{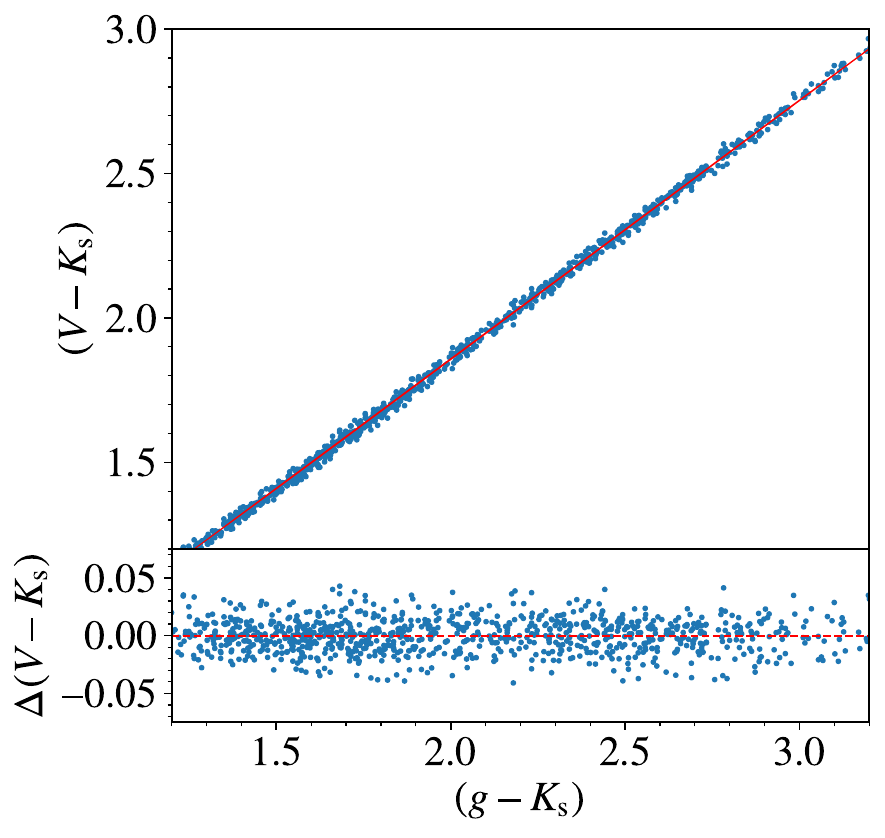}
\caption{The transformation between $(V-K_{\rm s})$ and $(g-K_{\rm s})$ colors derived from the calibrated sample of \citet{2013AJ....146...88C} after twice $3\sigma$ clipping. The best-fitting relation is shown as a red solid line, while the calibrated stars are represented by blue points.}
\label{fig:color_transformation}
\end{figure}

\setcounter{table}{0}
\renewcommand{\thetable}{A.\arabic{table}}
\begin{deluxetable}{cc}[!htbp]
\centering
\tabcolsep=1.5cm
\tablecaption{Coefficients of Calibration} \label{Coefficients of Calibration}
\tablehead{Coefficient & Value}
\startdata
$a_0$ & 0.14160 \\
$a_1$ & 0.77930 \\
$a_2$ & 0.05441 \\
$a_3$ & -0.0079 \\
\enddata
\end{deluxetable}
\FloatBarrier

\begin{deluxetable}{cccccccc}[!htbp]
\tabletypesize{\footnotesize}
\tabcolsep=15pt
\tablenum{A.2}
\tablecaption{
Photometric quantities and transformed colors used for the effective temperature determination.
\label{table:photometric_temperature}
}
\tablehead{
\colhead{ID} &
\colhead{$g$} &
\colhead{$K_{\rm s}$} &
\colhead{$(g-K_{\rm s})_0$} &
\colhead{$(V-K_{\rm s})_0$} &
\colhead{$T_{\rm eff}$} &
\colhead{$\sigma_{T_{\rm eff}}$}
}
\startdata
J0858-0809 & 10.539 & 7.737 & 2.719 & 2.504 & 4627 & 40 \\
J1930-1739 & 12.768 & 9.767 & 2.678 & 2.467 & 4638 & 37 \\
J1944-0837 & 13.040 & 9.799 & 2.662 & 2.453 & 4671 & 38 \\
J1950-1255 & 12.800 & 9.474 & 2.950 & 2.712 & 4500 & 37 \\
J1952-1836 & 12.204 & 8.651 & 3.082 & 2.830 & 4437 & 35 \\
J2001-1215 & 12.227 & 8.684 & 2.881 & 2.650 & 4516 & 36 \\
J2020-1002 & 12.732 & 9.613 & 2.925 & 2.689 & 4491 & 38 \\
J2023-0815 & 12.797 & 10.530 & 2.119 & 1.963 & 5160 & 50 \\
J2026-0047 & 12.492 & 9.817 & 2.139 & 1.981 & 5144 & 48 \\
J2038-0023 & 12.966 & 10.063 & 2.694 & 2.482 & 4647 & 37 \\
J2043-1305 & 12.089 & 9.885 & 2.096 & 1.942 & 5187 & 48 \\
J2140-1227 & 11.284 & 8.621 & 2.515 & 2.321 & 4761 & 38 \\
J2326-0159 & 11.400 & 8.308 & 2.947 & 2.709 & 4499 & 36 \\
\enddata
\tablecomments{
The listed colors are corrected for interstellar reddening using the extinction values adopted in Table~\ref{tab:observation_and_photometry_details}. 
Magnitudes and colors are given in mag, while $T_{\rm eff}$ and $\sigma_{T_{\rm eff}}$ are given in K. The effective temperatures listed here are the first-iteration input values derived from the transformed $(V-K_{\rm s})$ colors using the photometric metallicities in Table\,\ref{tab:observation_and_photometry_details}. The final adopted atmospheric parameters after the iterative analysis are listed in Table\,\ref{Table of Stellar Atmospheric Parameters}.
}
\end{deluxetable}
\FloatBarrier

\clearpage




\clearpage

\section{Na and K Resonance-Line Region of J2001-1215}

Figure~\ref{fig:NaK_region} shows the observed CFHT/ESPaDOnS spectrum of J2001-1215 around the Na\,{\sc i} D and K\,{\sc i} resonance-line regions. The Na\,{\sc i} D doublet is affected by interstellar absorption, preventing a reliable abundance determination. In the potassium region, the K\,{\sc i}\,$\lambda7664$ line is blended with strong telluric absorption, whereas the K\,{\sc i}\,$\lambda7698$ line lies between the major telluric absorption groups and remains sufficiently clean for abundance analysis.

\FloatBarrier
\setcounter{figure}{0}
\renewcommand{\thefigure}{B.\arabic{figure}}
\begin{figure*}[!htbp]
\centering
\includegraphics[width=1.0\linewidth]{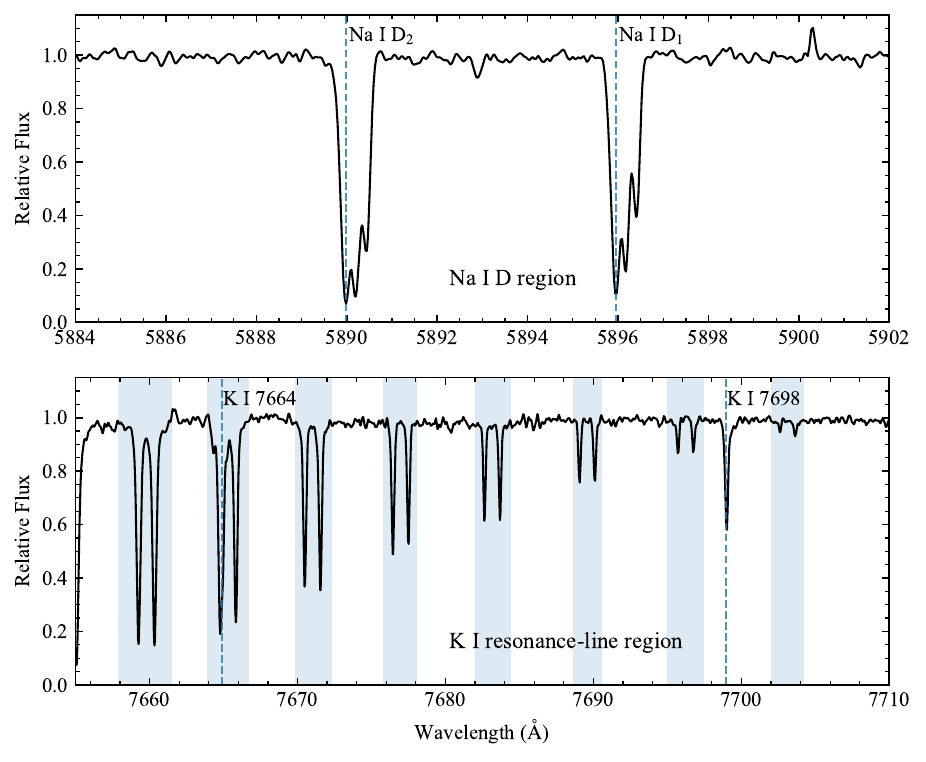}
\caption{Observed Na\,{\sc i} D and K\,{\sc i} resonance-line regions of J2001-1215. Upper panel: CFHT/ESPaDOnS spectrum around the Na\,{\sc i} D doublet. The Na\,{\sc i} D lines are affected by interstellar absorption, preventing a reliable sodium abundance determination. Lower panel: CFHT/ESPaDOnS spectrum around the K\,{\sc i} resonance-line region. The shaded regions indicate strong telluric absorption features. The K\,{\sc i}\,$\lambda7664$ line is blended with telluric absorption, whereas the K\,{\sc i}\,$\lambda7698$ line remains sufficiently free from contamination and is adopted for the potassium abundance determination.}
\label{fig:NaK_region}
\end{figure*}

\end{document}